\documentclass[11pt,a4paper]{article}

\usepackage{jcappub}
\usepackage{euscript,epsfig,amsmath,amssymb}
\usepackage{amsfonts,latexsym}

\usepackage{graphicx}

\usepackage{xcolor}
\usepackage{soul}

\begin{document}

\title{Axion mechanism of Sun luminosity: light shining through the solar radiation zone}

\author{V.D.~Rusov$^{1}$,}
\author{K.~Kudela$^{2}$,}
\author{I.V.~Sharph$^{1}$,}
\author{M.V.~Eingorn$^{3}$,}
\author{V.P.~Smolyar$^{1}$,}
\author{D.S.~Vlasenko$^{1}$,}
\author{T.N.~Zelentsova$^{1}$,}
\author{\\M.E.~Beglaryan$^{1}$,}
\author{and E.P.~Linnik$^{1}$}

\affiliation{$^{1}$Department of Theoretical and Experimental Nuclear Physics,\\ Odessa National Polytechnic University, Odessa, Ukraine}

\affiliation{$^{2}$Institute of Experimental Physics, SAS, Kosice, Slovakia}

\affiliation{$^{3}$CREST and NASA Research Centers, North Carolina Central University,\\ Durham, North Carolina, U.S.A.}

\emailAdd{siiis@te.net.ua (V.D. Rusov)}

\abstract{It is shown that the hypothesis of the axion mechanism of Sun luminosity suggesting that the solar axion particles are born in the core of the Sun
and may be efficiently converted back into $\gamma$-quanta in the magnetic field of the solar overshoot tachocline is physically relevant. As a result, it is
also shown that the intensity variations of the $\gamma$-quanta of axion origin, induced by the magnetic field variations in the tachocline,
directly cause the Sun luminosity %and total solar irradiance (TSI)
variations and eventually characterize the active and quiet states of the Sun.

Within the framework of this mechanism estimations of the strength of the axion coupling to a photon ($g_{a \gamma} = 3.6 \cdot 10^{-11} GeV^{-1}$) and the
hadronic axion particle mass ($m_a \sim 2.3 \cdot 10^{-2} eV$) have been obtained. It is also shown that the claimed axion parameters do not contradict any
known experimental and theoretical model-independent limitations.}

%\pacs{91.25.Cw,95.85.Ry,96.12.Hg,96.12.Wx}

\maketitle

\flushbottom

\

\section{Introduction}

A hypothetical pseudoscalar particle called axion is predicted by the theory related to solving the CP-invariance violation problem in QCD. The most important
parameter determining the axion properties is the energy scale $f_a$ of the so-called U(1) Peccei-Quinn symmetry violation. It determines both the axion mass
and the strength of its coupling to fermions and gauge bosons including photons. However, in spite of the numerous direct experiments, they have not been
discovered so far. Meanwhile, these experiments together with the astrophysical and cosmological limitations leave a rather narrow band for the permissible
parameters of invisible axion (e.g. $10^{-6} eV \leqslant m_a \leqslant 10^{-2} eV$~\cite{ref01,ref02}), which is also a well-motivated cold dark matter
candidate in this mass region \cite{ref01,ref02}.

A whole family of axion-like particles (ALP) with their own features may exist along with axions having the similar Lagrangian structure relative to the
Peccei-Quinn axion, as well as their own distinctive features. It consists in the fact that if they exist, the connection between their mass and their constant
of coupling to photons must be highly weakened, as opposed to the axions. It should be also mentioned that the phenomenon of photon-ALP mixing in the presence
of the electromagnetic field not only leads to the classic neutrino-like photon-ALP oscillations, but also causes the change in the polarization state of the
photons (the $a \gamma \gamma$ coupling acts like a polarimeter \cite{ref03}) propagating in the strong enough magnetic fields. It is generally assumed that
there are light ALPs coupled only to two photons, although the realistic models of ALPs with couplings both to photons and to matter are not excluded
\cite{ref04}. Anyway, they may be considered a well-motivated cold dark matter candidate \cite{ref01,ref02} under certain conditions, just like axions.

It is interesting to note that the photon-ALP mixing in magnetic fields of different astrophysical objects including active galaxies, clusters of galaxies,
intergalactic space and the Milky Way, may be the cause of the remarkable phenomena like dimming of stars luminosity (e.g., supernovae in the extragalactic
magnetic field \cite{ref06,ref07}) and "light  shining through a wall" (e.g., light from very distant objects, travelling through the Universe
\cite{ref03,ref05}). In the former case the luminosity of an astrophysical object is dimmed because some part of photons transforms into axions in the object's
magnetic field. In the latter case photons produced by the object are initially converted into axions in the object's magnetic field, and then after passing
some distance (the width of the "wall") are converted back into photons in another magnetic field (e.g., in the Milky Way), thus emulating the process of
effective free path growth for the photons in astrophysical medium \cite{ref08,ref09}.

For the sake of simplicity let us hereinafter refer to all such particles as axions if not stated otherwise.

In the present paper we consider the possible existence of the axion mechanism of Sun luminosity\footnote{Let us point out that the axion mechanism of Sun
luminosity used for estimating the axion mass was described for the first time in 1978 in the paper \cite{ref10}.} based on the "light shining through a wall"
effect. To be more exact, we attempt to explain the axion mechanism of Sun luminosity by the "light shining through a wall", when the photons born mainly in
the solar core are at first converted into axions via the Primakoff effect \cite{ref11} in its magnetic field, and then are converted back into photons after
passing the solar radiative zone and getting into the magnetic field of the overshoot tachocline. In addition to that we obtain the consistent estimates for
the axion mass ($m_a$) and the axion coupling constant to photons ($g_{a \gamma}$),
% nucleons (gan) and electrons ($g_{ae}$),
basing on this mechanism, and verify their values against the axion model results and the known experiments including CAST, ADMX, RBF.

\section{Photon-axion conversion and the case of maximal mixing}

Let us give some implications and extractions from the photon-axion oscillations theory which describes the process of the photon conversion into an axion and
back under the constant magnetic field $B$ of the length $L$. It is easy to show \cite{ref05,Raffelt-Stodolsky1988,ref07,Hochmuth2007} that in the case of the
negligible photon absorption coefficient ($\Gamma _{\gamma} \to 0$) and axions decay rate ($\Gamma _{a} \to 0$) the conversion probability is
\begin{equation}
P_{a \rightarrow \gamma} = \left( \Delta_{a \gamma}L \right)^2 \sin ^2 \left( \frac{ \Delta_{osc}L}{2} \right) \Big/ \left( \frac{ \Delta_{osc}L}{2}
\right)^2 \label{eq01}\, ,
\end{equation}
where the oscillation wavenumber $\Delta_{osc}$ is given by
\begin{equation}
\Delta_{osc}^2 = \left( \Delta_{pl} + \Delta_{Q,\perp} - \Delta_{a} \right)^2 + 4 \Delta_{a \gamma} ^2
\label{eq02}
\end{equation}
while the mixing parameter $\Delta _{a \gamma}$, the axion-mass parameter $\Delta_{a}$, the refraction parameter $\Delta_{pl}$ and the QED dispersion parameter
$\Delta_{Q,\perp}$ may be represented by the following expressions:
\begin{equation}
\Delta _{a \gamma} = \frac{g_{a \gamma} B}{2} = 540 \left( \frac{g_{a \gamma}}{10^{-10} GeV^{-1}} \right) \left( \frac{B}{1 G} \right) ~~ pc^{-1}\, ,
\label{eq03}
\end{equation}
\begin{equation}
\Delta _{a} = \frac{m_a^2}{2 E_a} = 7.8 \cdot 10^{-11} \left( \frac{m_a}{10^{-7} eV} \right)^2 \left( \frac{10^{19} eV}{E_a} \right) ~~ pc^{-1}\, ,
\label{eq04}
\end{equation}
\begin{equation}
\Delta _{pl} = \frac{\omega ^2 _{pl}}{2 E_a} = 1.1 \cdot 10^{-6} \left( \frac{n_e}{10^{11} cm^{-3}} \right) \left( \frac{10^{19} eV}{E_a} \right) ~~ pc^{-1}\,
, \label{eq05}
\end{equation}
\begin{equation}
\Delta _{Q,\perp} = \frac{m_{\gamma, \perp}^2}{2 E_a}\, . \label{eq06}
\end{equation}

Here $g_{a \gamma}$ is the constant of axion coupling to photons; $B$ is the transverse magnetic field; $m_a$ and $E_a$ are the axion mass and energy; $\omega
^2 _{pl} = 4 \pi \alpha n_e / m_e$ is an effective photon mass in terms of the plasma frequency if the process does not take place in vacuum, $n_e$ is the
electron density, $\alpha$ is the fine-structure constant, $m_e$ is the electron mass; $m_{\gamma, \perp}^2$ is the effective mass square of the transverse
photon which arises due to interaction with the external magnetic field.

The conversion probability (\ref{eq01}) is energy-independent, when $2 \Delta _{a \gamma} \approx \Delta_{osc}$, i.e.
\begin{equation}
P_{a \rightarrow \gamma} \cong \sin^2 \left( \Delta _{a \gamma} L \right)\, ,
\label{eq07}
\end{equation}
or, whenever the oscillatory term in (\ref{eq01}) is small ($\Delta_{osc} L / 2 \to 0$), implying the limiting coherent behavior
\begin{equation}
P_{a \rightarrow \gamma} \cong \left( \frac{g_{a \gamma} B L}{2} \right)^2\, .
\label{eq08}
\end{equation}

%As it is shown in \cite{ref05}, in the energy-independent case (\ref{eq07}), when the following inequalities are satisfied:
%\begin{equation}
%\Delta_{a \gamma} \gg \Delta _{Q,\perp}, ~~~ 2 \Delta_{a \gamma} \gg \Delta_a, ~~~ 2 \Delta_{a \gamma} \gg \Delta_{pl}\, , \label{eq09}
%\end{equation}
%it is easy to derive the following relations by means of (\ref{eq03})-(\ref{eq06}):
%\begin{equation}
%\left( \frac{B}{1 G} \right) \left( \frac{E_a}{10^{19} eV} \right) \ll 7.52 \cdot 10^{-9}\, , \label{eq10}
%\end{equation}
%\begin{equation}
%E_a \gg 70 eV \left( \frac{m_a}{10^{-9} eV} \right)^2  \left( \frac{B}{1 G} \right)^{-1} \left( \frac{10^{-10} GeV^{-1}}{g_{a \gamma}} \right)\, , \label{eq11}
%\end{equation}
%\begin{equation}
%n_e \ll 10^{20} cm^{-3} \left( \frac{E_a}{10^{19} eV} \right)  \left( \frac{B}{1 G} \right)\, . \label{eq12}
%\end{equation}

%The condition of maximum mixing (i.e. $P_{a \rightarrow \gamma} \to 1$ in (\ref{eq07})) \hl{is obviously provided by the size $L$ of the region
%that satisfies the conditions (\ref{eq10})-(\ref{eq12}) together with the following expression:
%equal to the oscillation length, which is}

It is worth noting that the oscillation length corresponding to (\ref{eq07}) reads
\begin{equation}
L_{osc} = \frac{\pi}{\Delta_{a \gamma}} = \frac{2 \pi}{g_{a \gamma} B} \cong 5.8 \cdot 10^{-3}
\left( \frac{10^{-10} GeV^{-1}}{g_{a \gamma}} \right)
\left( \frac{1G}{B} \right)  ~pc
\label{eq13}
\end{equation}
%taking into account that $\Delta_{osc} \approx 2 \Delta_{a \gamma}$.
\noindent assuming a purely transverse field. In the case of the appropriate size $L$ of the region a complete transition between photons and axions is
possible.

From now on we are going to be interested in the energy-independent case (\ref{eq07})
%and the relations (\ref{eq10})-(\ref{eq13})
or (\ref{eq08}) which play the key role in determination of the axion mechanism of Sun luminosity hypothesis parameters (the axion coupling constant to
photons $g_{a \gamma}$, the transverse magnetic field $B$ of length $L$ and the axion mass $m_a$).

\

\section{Axion mechanism of Sun luminosity}

Our hypothesis is that the solar axions which are born in the solar core \cite{ref01,ref02} through the known Primakoff effect \cite{ref11}, may be converted
back into $\gamma$-quanta in the magnetic field of the solar tachocline (the base of the solar convective zone). The magnetic field variations in the
tachocline cause the converted $\gamma$-quanta intensity variations in this case, which in their turn cause the variations of the Sun luminosity known as the
active and quiet Sun states. Let us consider this phenomenon in more detail below.

As we noted above, the expression (\ref{eq01}) for the probability of the axion-photon oscillations in the transversal magnetic field was obtained for the
media with the quasi-zero refraction, i.e. for the media with a negligible photon absorption coefficient ($\Gamma_{\gamma} \to 0$). It means that in order for
the axion-photon oscillations to take place without any significant losses, a medium with a very low or quasi-zero density is required, which would suppress
the processes of photon absorption almost entirely.

Surprisingly enough, it turns out that such "transparent" media can take place, and not only in plasmas in general, but straight in the convective zone of the
Sun. Here we generally mean the so-called magnetic flux tubes, the properties of which are examined below.

\

\subsection{Channeling of $\gamma$-quanta along the magnetic flux tubes (waveguides) in the Sun convective zone}

\label{subsec-channeling}

\begin{figure*}
%\centerline{\includegraphics[width=12cm]{TachoclineFluxTubes-3.eps}}
\centerline{\includegraphics[width=12cm]{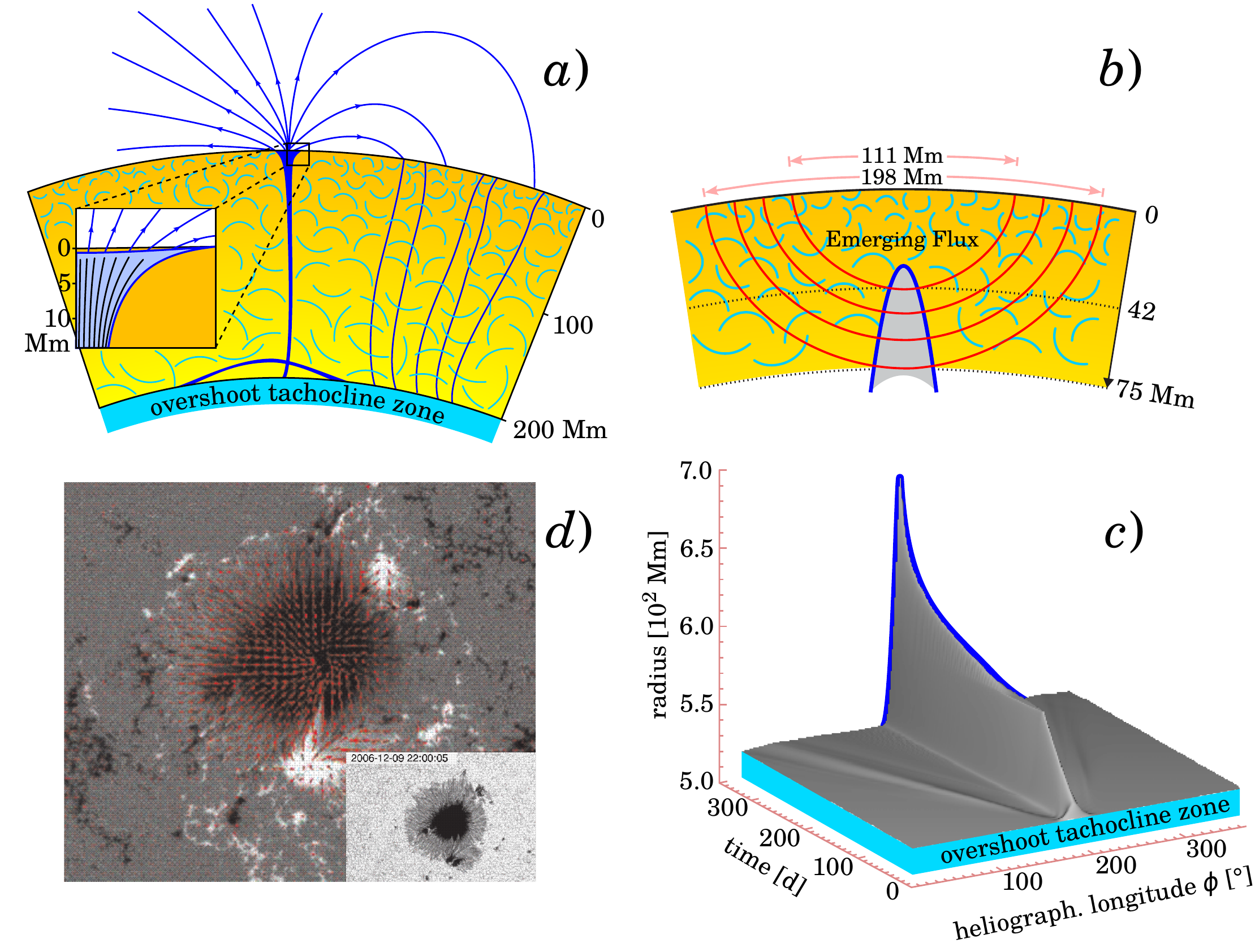}}
 \caption{(a) Vertical cut through an active region illustrating the connection between a
sunspot at the surface and its origins in the toroidal field layer at the base of the convection zone. Horizontal fields stored at the base of the convection
zone (the overshoot tachocline zone) during the cycle. Active regions form from sections brought up by buoyancy (one shown in the process of rising). After
eruption through the solar surface a nearly potential field is set up in the atmosphere (broken lines), connecting to the base of the convective zone via
almost vertical flux tube. Hypothetical small scale structure of a sunspot is shown in the inset (Adopted from Spruit~\cite{ref18} and Spruit and
Roberts~\cite{ref15}). \newline (b) Detection of emerging sunspot regions in the solar interior~\cite{ref18}. Acoustic ray paths with lower turning points
between 42 and 75 Mm (1 Mm=1000 km) crossing a region of emerging flux. For simplicity, only four out of total of 31 ray paths used in this study (the
time-distance helioseismology experiment) are shown here. Adopted from~\cite{ref19}. \newline (c) Emerging and anchoring of stable flux tubes in the overshoot
tachocline zone, and its time-evolution in the convective zone. Adopted from \cite{ref20}. \newline (d) Vector magnetogram of the white light image of a
sunspot (taken with SOT on a board of the Hinode satellite -- see inset) showing in red the direction of the magnetic field and its strength (length of the
bar). The movie shows the evolution in the photospheric fields that has led to an X class flare in the lower part of the active region. Adopted
from~\cite{ref21}.\label{fig01}}
\end{figure*}

The idea of the energy flow channeling along a fanning magnetic field has been suggested for the first time by Hoyle~\cite{ref12} as an explanation for
darkness of umbra of sunspots. It was incorporated in a simple sunspot model by Chitre~\cite{ref13}. Zwaan~\cite{ref14} extended this suggestion to smaller
flux tubes to explain the dark pores and the bright faculae as well. Summarizing the research of the convective zone magnetic fields in the form of the
isolated flux tubes, Spruit and Roberts~\cite{ref15} suggested a simple mathematical model for the behavior of thin magnetic flux tubes, dealing with the
nature of the solar cycle, sunspot structure, the origin of spicules and the source of mechanical heating in the solar atmosphere. In this model, the so-called
thin tube approximation is used (see \cite{ref15} and references therein), i.e. the field is conceived to exist in the form of slender bundles of field lines
(flux tubes) embedded in a field-free fluid (Fig.~\ref{fig01}). Mechanical equilibrium between the tube and its surrounding is ensured by a reduction of the
gas pressure inside the tube, which compensates the force exerted by the magnetic field.  In our opinion, this is exactly the kind of mechanism
Parker~\cite{ref16} was thinking about when he wrote about the problem of flux emergence: "Once the field has been amplified by the dynamo, it needs to be
released into the convection zone by some mechanism, where it can be transported to the surface by magnetic buoyancy"~\cite{ref17}.

In order to understand magnetic buoyancy, let us consider an isolated horizontal flux tube in pressure equilibrium with its non-magnetic surroundings
(Fig.~\ref{fig01}). The magnetic field $\vec{B}$ alternating along the vertical axis $z$ induces the vortex electric field in the magnetic flux tube containing
dense plasma. The charged particles rotation in plasma with the angular velocity $\omega$ leads to a centrifugal force per unit volume
\begin{equation}
\vert \vec{F} \vert \sim \rho \vert \vec{\omega} \vert ^2 r\, . \label{eq14}
\end{equation}

If we consider this problem in a rotating noninertial reference frame, such noninertiality, according to the equivalence principle, is equivalent to
"introducing" a radial non-uniform gravitational field with the "free fall acceleration"
\begin{equation}
g (r) = \vert \vec{\omega} \vert ^2 r\, . \label{eq15}
\end{equation}

\begin{figure*}
%\centerline{\includegraphics[width=9cm]{hydrostatic_equilibrium.eps}}
\centerline{\includegraphics[width=9cm]{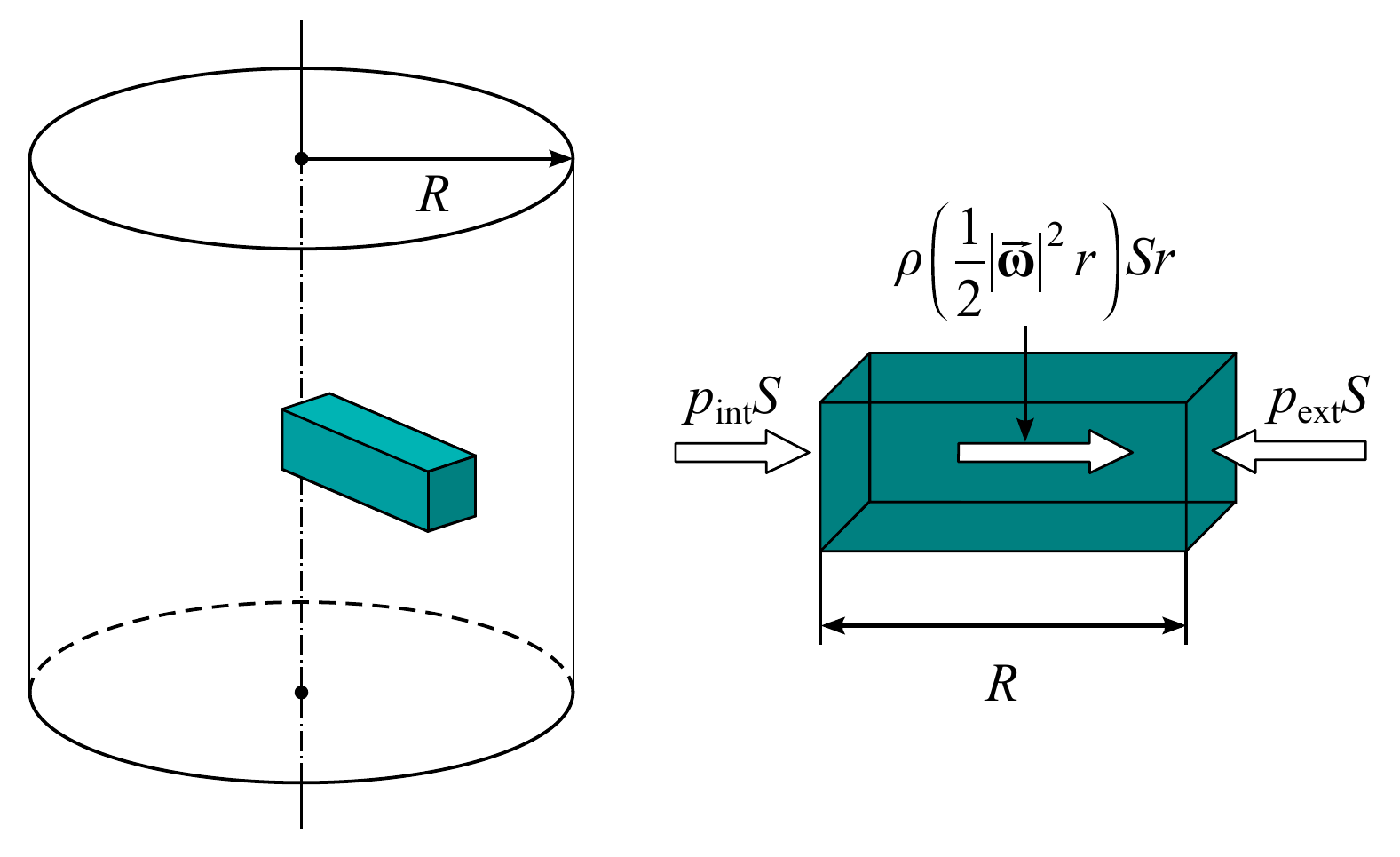}}

\caption{Representation of a "hydrostatic equilibrium" in the rotating "liquid" of a
magnetic flux tube. \label{fig02}}
\end{figure*}

In such case the pressure difference inside ($p_{int}$) and outside ($p_{ext}$) the rotating "liquid" of the tube may be treated by analogy with the regular
hydrostatic pressure (Fig.~\ref{fig02}).

Let us pick a radial "liquid column" inside the tube as it is shown in Fig.~\ref{fig02}. Since the "gravity" is non-uniform in this column, it is equivalent to
a uniform field with "free fall acceleration"
\begin{equation}
\left \langle g(r) \right \rangle = \frac{1}{2} \vert \vec{\omega} \vert ^2 R\, , \label{eq16}
\end{equation}
where $R$ is the tube radius which plays a role of the "liquid column height" in our analogy.

By equating the forces acting on the chosen column similar to hydrostatic pressure (Fig.~\ref{fig02}), we derive:
\begin{equation}
p_{ext} = p_{int} + \frac{1}{2} \rho \vert \vec{\omega} \vert^2 R^2\, . \label{eq17}
\end{equation}

This raises the question as to what physics is hidden behind the "centrifugal" pressure. In this relation, let us consider the magnetic field energy density
\begin{equation}
w_B = \frac{\vert \vec{B} \vert ^2}{2 \mu _0}\, , \label{eq18}
\end{equation}
where $\mu_0$ is the magnetic permeability of vacuum.

Suppose that the total magnetic field energy of the "growing" tube grows linearly between the tachocline and the photosphere. In this case if the average total
energy of the magnetic field in the tube transforms into the kinetic energy of the tube matter rotation completely, it is easy to show that
\begin{equation}
\left \langle E_B \right \rangle = \frac{1}{2} w_B V = \frac{1}{2} \frac{\vert \vec{B} \vert ^2}{2 \mu_0} V = \frac{ I \vert \vec{\omega} \vert^2}{2}\, ,
\label{eq19}
\end{equation}
where $I = mR^2 / 2$ is the tube's moment of inertia about the rotation axis, $m$ and $V$ are the mass and the volume of the tube medium respectively.

From (\ref{eq19}) it follows that
\begin{equation}
\frac{\rho \vert \vec{\omega} \vert ^2 R^2}{2} = \frac{\vert \vec{B} \vert ^2}{2 \mu_0}\, . \label{eq20}
\end{equation}

Finally, substituting (\ref{eq20}) into (\ref{eq17}) we obtain the desired relation
\begin{equation}
p_{ext} = p_{int} + \frac{\vert \vec{B} \vert^2}{2 \mu_0} \, ,
\label{eq21}
\end{equation}
which is exactly equal to the well-known expression by Parker~\cite{ref22}, describing the so-called self-confinement of force-free magnetic fields.

In spite of the obvious, though turned out to be surmountable, difficulties of the expression (\ref{eq14}) application to the real problems, it was shown
(see~\cite{ref15} and Refs. therein) that strong buoyancy forces act in magnetic flux tubes of the required field strength ($10^4 - 10^5 ~G$~\cite{ref23}).
Under their influence tubes either float to the surface as a whole (e.g., Fig.~1 in \cite{ref24}) or they form loops of which the tops break through the
surface (e.g., Fig.~1 in~\cite{ref14}) and lower parts descend to the bottom of the convective zone, i.e. to the overshoot tachocline zone. The convective
zone, being unstable, enhanced this process~\cite{ref25,ref26}. Small tubes take longer to erupt through the surface because they feel stronger drag forces.
It is interesting to note here that the phenomenon of the drag force which
raises the magnetic flux tubes to the convective surface with the speeds about
0.3-0.6~km/s, was discovered in direct experiments using the method of time-distance helioseismology~\cite{ref19}. Detailed calculations of the
process~\cite{ref27} show that even a tube with the size of a very small spot, if located within the convective zone, will erupt in less than two years. Yet,
according to~\cite{ref27}, horizontal fields are needed in overshoot tachocline zone, which survive for about 11~yr, in order to produce an activity cycle.

In other words, a simplified scenario of magnetic flux tubes (MFT) birth and space-time evolution (Fig.~\ref{fig01}a) may be presented as follows. MFT is born
in the overshoot tachocline zone (Fig.~\ref{fig01}c) and rises up to the convective zone surface (Fig.~\ref{fig01}b) without separation from the tachocline
(the anchoring effect), where it forms the sunspot (Fig.~\ref{fig01}d) or other kinds of active solar regions when intersecting the photosphere.  There are
more fine details of MFT physics expounded in overviews by Hassan \cite{ref17} and Fisher \cite{ref24}, where certain fundamental questions, which need to be
addressed to understand the basic nature of magnetic activity, are discussed in detail: How is the magnetic field generated, maintained and dispersed? What are
its properties such as structure, strength, geometry? What are the dynamical processes associated with magnetic fields? What role do magnetic fields play in
energy transport?

Dwelling on the last extremely important question associated with the energy transport, let us note that it is known that thin magnetic flux tubes can support
longitudinal (also called sausage), transverse (also called kink), torsional (also called torsional Alfv\'{e}n), and fluting modes
(e.g.,~\cite{ref28,ref29,ref30,ref31,ref32}); for the tube modes supported by wide magnetic flux tubes, see Roberts and Ulmschneider~\cite{ref31}. Focusing on
the longitudinal tube waves known to be an important heating agent of solar magnetic regions, it is necessary to mention the recent papers by
Fawzy~\cite{ref33}, which showed that the longitudinal flux tube waves are identified as insufficient to heat the solar transition region and corona in
agreement with previous studies~\cite{ref34}.

In other words, the problem of generation and transport of energy by magnetic flux tubes remains unsolved in spite of its key role in physics of various types
of solar active regions.

Interestingly, this problem may be solved in a natural way in the framework of the "axion" model of the Sun. It may be shown that the inner pressure,
temperature and matter density decrease rapidly in a magnetic tube "growing" between the tachocline and the photosphere. This fact is very important, since the
rapid decrease of these parameters predetermines the decrease of the medium opacity inside the magnetic tube, and, consequently, increases the photon free path
inside the magnetic tube drastically. In other words, such decrease of the pressure, temperature and density inside the "growing" magnetic flux tube is a
necessary condition for the virtually ideal (i.e. without absorption) photon channeling along such magnetic tube. The latter specifies the axion-photon
oscillations efficiency inside the practically empty magnetic flux tubes.

\

\subsection{Hydrostatic equilibrium and a sharp tube medium cooling effect}

Let us assume that the tube has the length $l(t)$ by the time $t$. Then its volume is equal to $S \cdot l(t)$ and the heat capacity is
\begin{equation}
c \cdot \rho \cdot S l(t)\, , \label{eq22}
\end{equation}
where $S$ is the tube cross-section, $\rho (t)$ is the density inside the tube, $c$ is the specific heat capacity.

If the tube becomes longer by $ v(t)dt$ for the time $dt$, then the magnetic field energy increases by
\begin{equation}
\frac{1}{2} \frac{\vert \vec{B} \vert ^2}{2 \mu_0} S \cdot v(t) dt\, , \label{eq23}
\end{equation}
where $v(t)$ is the tube propagation speed.

The matter inside the tube, obviously, has to cool by the temperature $dT$ so that the internal energy release maintained the magnetic energy growth.
Therefore, the following equality must hold:
\begin{equation}
c \rho(t) l(t) \frac{dT}{dt} S = - \frac{1}{2} \frac{\vert \vec{B} \vert ^2}{2 \mu_0} S \cdot v(t)\, . \label{eq24}
\end{equation}

Taking into account the fact that the tube grows practically linearly~\cite{ref19}, i.e. $vt = l$, Parker relation (\ref{eq21}) and the tube's equation of state
\begin{equation}
p_{int} (t) = \frac{\rho}{\mu_{*}} R_{*} T(t) \Leftrightarrow \rho = \frac{\mu_{*}}{R_{*}} \frac{p_{int}}{T(t)}\, ,
\end{equation}
the equality (\ref{eq24}) may be rewritten (by separation of variables) as follows:
\begin{equation}
\frac{dT}{T} = - \frac{R_{*}}{2 c \mu_{*}} \left[ \frac{p_{ext}}{p_{int} (t)} - 1 \right] \frac{dt}{t}\, , \label{eq25}
\end{equation}
where $\mu_{*}$ is the tube matter molar mass, $R_{*}$ is the universal gas constant.

After integration of (\ref{eq25}) we obtain
\begin{equation}
\ln \left[ \frac{T(t)}{T(0)} \right] = - \frac{R_{*}}{2 c \mu_{*}} \int \limits_0 ^t \left[ \frac{p_{ext}}{p_{int} (\tau)} -1 \right] \frac{d\tau}{\tau}\, .
\label{eq26}
\end{equation}

It is easy to see that the multiplier $(1/\tau)$ in (\ref{eq26}) assigns the region near $\tau = 0$ in the integral. The integral converges, since
\begin{equation}
\lim \limits_{\tau \to 0} \left[ \frac{p_{ext}}{p_{int} (\tau)} -1 \right] = 0\, .
\end{equation}

Expanding $p_{int}$ into Taylor series and taking into account that $p_{int}(\tau=0) = p_{ext}$, we obtain
\begin{equation}
p_{int} (\tau) = p_{ext} + \frac{dp_{int} (\tau = 0)}{d\tau}\tau = p_{ext} (1 - \gamma \tau)\, , \label{eq27}
\end{equation}
\noindent where
\begin{equation}
\gamma = - \frac{1}{p_{ext}} \frac{dp_{int} (\tau = 0)}{d\tau} = - \frac{1}{p_{int} (\tau = 0)} \frac{dp_{int} (\tau = 0)}{d\tau}\, . \label{eq28}
\end{equation}

From (\ref{eq27}) it follows that
\begin{equation}
\frac{p_{ext}}{p_{int}(\tau)} - 1 = \frac{1}{1 - \gamma \tau} - 1 \approx \gamma \tau\, , \label{eq29}
\end{equation}
therefore, substituting (\ref{eq29}) into (\ref{eq26}), we find its solution in the form
\begin{equation}
T(t) = T(0) \exp \left( - \frac{R_{*}}{2 c \mu_{*}} \gamma t \right)\, . \label{eq30}
\end{equation}

It is extremely important to note here that the solution (\ref{eq30}) points at the remarkable fact that at least at the initial stages of the tube formation
its temperature decreases exponentially, i.e. very sharply. The same conclusion can be made for the pressure and the matter density in the tube. Let us show
it.

Assuming that relation (\ref{eq28}) holds not only for $\tau = 0$, but also for small $\tau$ close to zero, we obtain
\begin{equation}
p_{int}(t) = p_{int} (0) \exp (-\gamma t)\, , \label{eq31}
\end{equation}
i.e. the inner pressure also decreases exponentially, but with the different exponential factor.

Further, assuming that the heat capacity of a molecule in the tube is
\begin{equation}
c = \frac{i}{2} \frac{R_{*}}{\mu_{*}}\, , \label{eq32}
\end{equation}
where $i$ is a number of molecule's degrees of freedom, and substituting the expressions (\ref{eq30}) and (\ref{eq31}) into the tube's equation of state, we
derive the expression for the matter density in the tube
\begin{equation}
\rho (t) = \frac{\mu_{*}}{R_{*}} \frac{p_{int} (0)}{T(0)} \exp \left[ - \left( 1 - \frac{1}{i} \right) \gamma t \right]\, . \label{eq33}
\end{equation}

Taking into account that $i \geqslant 3$, we can see that the density decreases exponentially just like the temperature (\ref{eq30}) and the pressure
(\ref{eq31}).

\

\subsection{Ideal photon channeling (without absorption) conditions inside the magnetic flux tubes}
\label{subsec-ideal-channeling}

%Let us calculate the inner pressure $p_{int}$ of the magnetic flux tube in the solar tachocline. In the framework of the standard %model of the Sun the pressure
%in the tachocline zone is about $\sim 6 \cdot 10^{12} ~Pa$, while the magnetic field strength reaches 400 T, according to our %estimates (Fig.~\ref{fig03}).

% Initial place for Fig.03

%Then from the hydrostatic condition by Parker (\ref{eq21}) it follows that the inner pressure of the magnetic flux tube in the solar %tachocline is equal to
%\begin{equation}
%p_{int} \sim 6 \cdot 10^{12} - 4 \cdot 10^{10} \cong 6 \cdot 10^{12} ~[Pa],
%\label{eq34}
%\end{equation}

%\noindent which is comparable to the external pressure.

%However, the situation changes drastically

It should be stressed that there is no thermal equilibrium between the magnetic tube and its surroundings: according to (\ref{eq30}), (\ref{eq31}) and
(\ref{eq33}), the temperature and the matter density decrease together with the inner pressure, and the decrease is sharply exponential, because the
exponential factor in (\ref{eq29}) becomes very large
\begin{equation}
\gamma \tau = \frac{p_{ext}}{p_{int} (\tau)} - 1 \to \infty\, . \label{eq36}
\end{equation}

\begin{figure*}
\noindent
\begin{center}
\includegraphics[width=7cm]{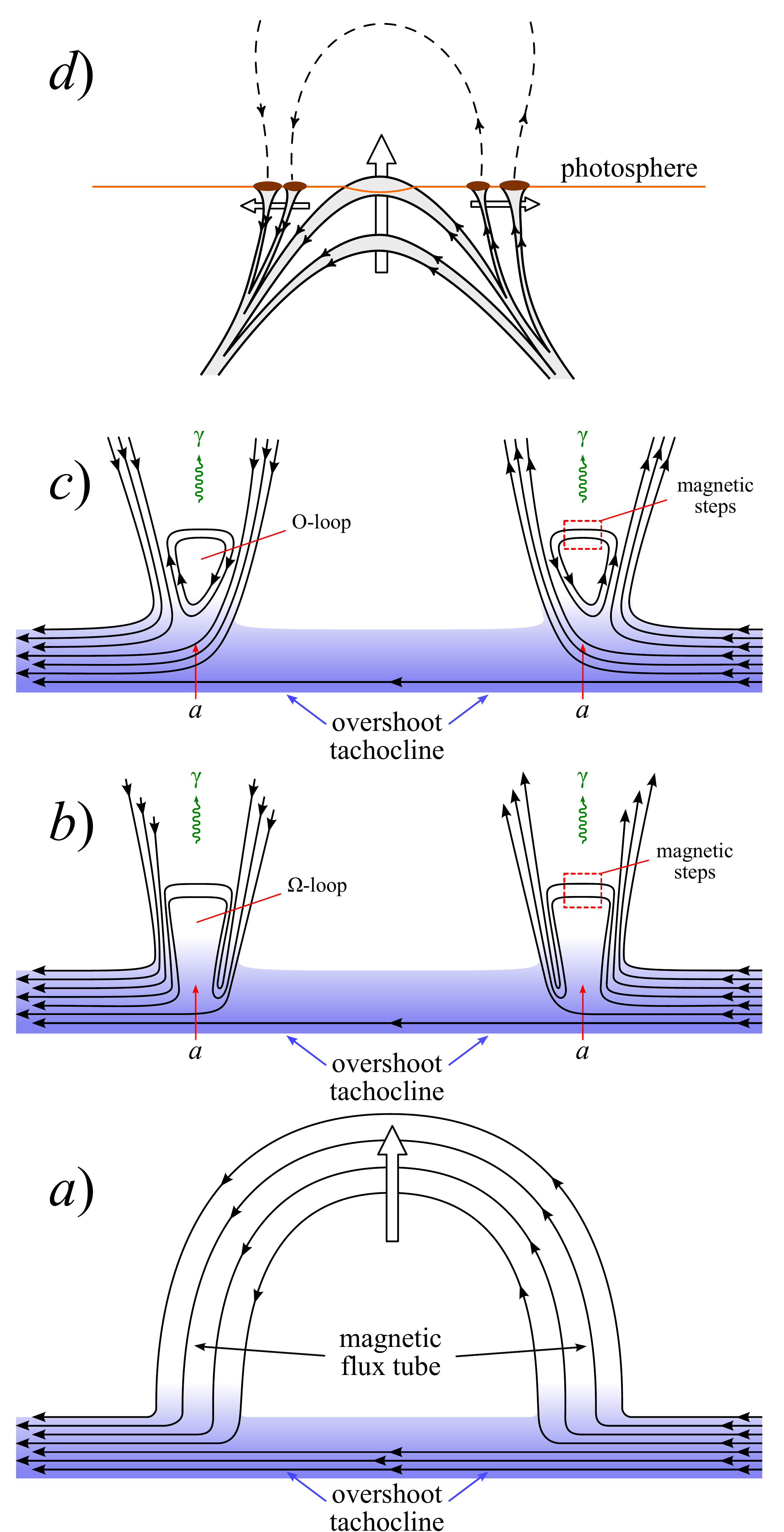}
\end{center}

%\centerline{\includegraphics[width=10cm]{tachocline-magnetic_tube-7.pdf}}

\caption{(a) Magnetic loop tubes (black lines) formation in the tachocline
               through the shear flows instability development. \newline
               (b)
               % "Capillary" effect
               Topological features of the $\Omega$-loop magnetic
               configurations inside the magnetic tube and the sketch of the
               axions (red arrows) conversion into $\gamma$-quanta (green arrows)
               inside the magnetic flux tubes containing the magnetic steps.
               The buoyancy of the magnetic steps is limited by the magnetic
               tube interior cooling.
               %Here $L_{MS}$} is the height of the magnetic shear steps.
               The tubes' rotation is not shown here for
               the sake of simplicity. \newline
               %\colorbox{yellow}{\parbox{\linewidth}{
               (c) Consequent topological effects of the magnetic reconnection
               in the magnetic tubes (see~\cite{Priest2000}), where the $\Omega$-loop
               reconnects across its base, pinching off the $\Omega$-loop to from
               a free O-loop (see Fig.~4 in~\cite{Parker1994}). The buoyancy
               of the O-loop is also limited by the magnetic tube interior cooling.
               %}}
               \newline
               (d) Emergence of magnetic flux bundle and coalescence of spots
               to explain the phenomenology of active region emergence (adopted
               from Zwaan~\cite{ref14}; Spruit~\cite{ref18}).}
\label{fig04}
\end{figure*}

\begin{figure*}
%\centerline{\includegraphics[width=15cm]{ShearStability-Yohkoh-4.eps}}
\centerline{\includegraphics[width=15cm]{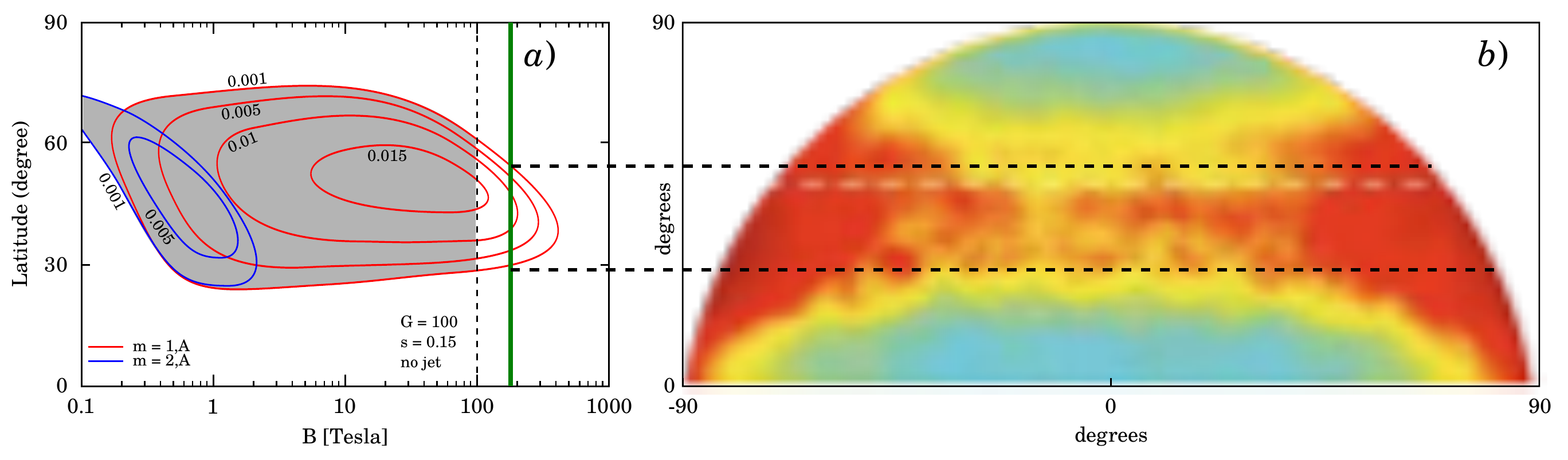}}

\caption{(a) Growth rates for magnetic shear instabilities are plotted as functions of the initial latitude (vertical axes) and the field strength (horizontal
axes) of a toroidal band.  Shaded areas indicate instability in 0.1-100~T band (gray) and $\sim$~200~T band (green). Contour lines represent $m = 1$ and $m =
2$ antisymmetric (A) modes as indicated. The non-dimensional model is normalized in such a way that the growth rate of 0.01 corresponds to an e-folding growth
time of 1 year.  The parameter s is the fractional angular velocity contrast between equator and pole and the reduced gravity G (adopted from \cite{ref35}). In
addition, a hidden part of the "latitude -- magnetic field in overshoot tachocline zone" dependence, which was missing on the original plot (Fig.~11
in~\cite{ref35}), is plotted to the right of the dashed line. \newline (b) Solar images at photon energies from 250 eV up to a few keV from the Japanese X-ray
telescope Yohkoh (1991-2001) (adopted from~\cite{ref36}). The following shows solar X-ray activity during the last maximum of the 11-year solar cycle.
\label{fig03}}
\end{figure*}

Because of the fact that the pressure, density and temperature in the tube (which under the conditions of nonthermal equilibrium (see also \cite{ref22,ref83})
grows from tachocline to photosphere during $1\, -\, 2$ years \cite{ref19}) are ultralow, they virtually do not influence the radiation transport in these
tubes at all. In other words, the Rosseland free paths for $\gamma$-quanta inside the tubes are so big that the Rosseland
opacity~\cite{ref37,ref38,ref39,ref40,ref40a,ref41,ref42,ref43} and the absorption coefficients ($\Gamma _{\gamma} \to 0$) tend to zero. The low refractivity
(or high transparency) is achieved for the limiting condition (following from (\ref{eq21}) for the ultralow inner pressure):
\begin{equation}
p_{ext} = \frac{\vert \vec{B} \vert ^2}{2 \mu_0}\, ,
\label{eq37}
\end{equation}
%in complete agreement with the argumentation in~\cite{ref83}. Furthermore, within the framework of the "axion" model of the Sun, since the overshoot tachocline
%zone is located substantially higher than the base of the convective zone (see Fig.~\ref{fig04} and the corresponding text) in this model, let us suppose that
%the plasma pressure outside the magnetic tube near the new overshoot tachocline zone location falls by an order of magnitude and is about $\sim 10^{11}~ Pa$.
%In such case it is easy to see that the overshoot tachocline magnetic field $\sim 400 ~T$ (see Fig.~\ref{fig03}) compensates the outer pressure almost
%completely, so the condition (\ref{eq37}) really takes place and the inner pressure is ultralow ($p_{int} \to 0 ~[Pa]$).
\noindent in complete agreement with the argumentation e.g. in~\cite{ref83}. Furthermore, within the framework of the "axion" model of the Sun, the pressure of
plasma outside the magnetic tube, starting slightly above the overshoot tachocline, grows and makes about $10^{13}~ Pa$. It is easy to see that in this case
the magnetic field inside the tube (Fig.~\ref{fig04}b,c) is substantially higher than that of the overshoot tachocline ($\sim 200 ~T$ (see Fig.~\ref{fig03})),
and compensates the outer pressure almost completely, so that the condition (\ref{eq37}) holds true and the inner gas pressure is ultralow ($p_{int} \to 0
~[Pa]$).

%In addition, it is not difficult to show that the formula (\ref{eq37}) holds true for the photosphere surface as well,
%where the magnetic field $\sim 1600 ~G$ \cite{ref84}.

The obtained results validate the substantiation of an almost ideal photon 
channeling mechanism along the magnetic flux tubes. According to current 
understanding of the magnetic tubes formation processes in the tachocline, the 
following scenario may be supposed in the framework of the axion mechanism of
solar luminosity. At the first stage this process is determined by the numerous
magnetic tubes appearance (Fig.~\ref{fig04}a) as a consequence of the shear 
flow instability development in the tachocline. As is shown above (e.g. 
(\ref{eq31})), the pressure inside these tubes becomes ultralow causing the 
formation of the topological features of the magnetic field, e.g. the "magnetic
steps" in the cool part of the tube ($\Omega$-loops in Fig.~\ref{fig04}b), 
which makes the formation of the current sheet and, consequently, realization 
of the reconnection processes in the magnetic tube (O-loops in 
Fig.~\ref{fig04}c) possible.
%and rise
%of the magnetic steps in the tubes (Fig.~\ref{fig04}b).To put it differently, a
%kind of the magnetic capillary effect is observed in this case.
A complete picture of the magnetic tube's space-time evolution in the convective zone shown in Fig.~\ref{fig04} is an illustration of the axion to
$\gamma$-quanta conversion process (within the "magnetic steps" limited by the buoyancy drop in the cool region of the tube (Fig.~\ref{fig04}b,c)) with
the formation of an active solar region in the photosphere (Fig.~\ref{fig04}d). It thereby reveals the nature of the unique energy transport mechanism through
the convective zone by magnetic flux tubes.

%\noindent \colorbox{yellow}{\parbox{\linewidth}{
Although there is currently no deep and detailed understanding of the process of buoyant magnetic tubes formation in the tachocline
\cite{ref34-3,ref35-3,ref36-3,ref37-3,ref38-3,ref39-3,ref40-3}, the axion mechanism of Sun luminosity, which determines the process of $\gamma$-quanta
channeling along the magnetic tubes in the convection zone, gives grounds for the following presumable picture of their formation.

\begin{figure}
%\noindent \colorbox{yellow}{\parbox{\linewidth}{
\begin{center}
\includegraphics[width=10cm]{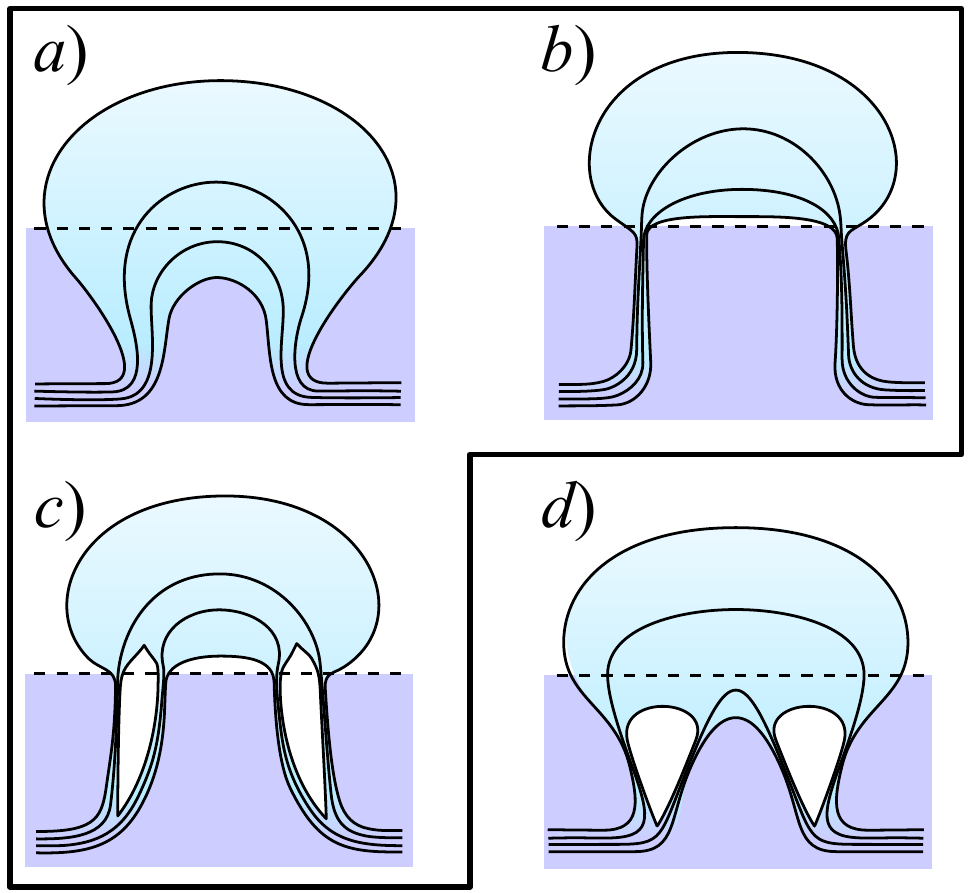}
\end{center}
\caption{The possible ways of a toroidal magnetic flux tube development into a sunspot.\newline (a) A rough representation of the form a tube can take after
the rise to the surface by magnetic buoyancy (adopted from Fig.~2a in \cite{ref41-3});\newline (b) demonstrates the "crowding" below the photosphere surface
because of cooling (adopted from Fig.~2b in \cite{ref41-3});\newline (c) demonstrates the tube splitting as a consequence of the sharp narrowing above the cool
region (adopted from Fig.~2c in \cite{ref41-3});\newline (d) shows the splitting of the magnetic tube because of the inner region cooling under the conditions
when the tube is in the thermal disequilibrium with its surroundings and the convective heat transfer is suppressed \cite{ref42-3} below $\sim 0.85 R_{Sun}$.
The transition from the cool region to the quickly expanding hot region above $\sim 0.85 R_{Sun}$, where the tube is in the thermal equilibrium with its
surroundings, causes and explains the appearance of the neutral atoms in the upper convection zone (in contrast to Fig.~2c in~\cite{ref41-3}). }
\label{fig04-3}
%}}
\end{figure}

First, we suppose that the classical mechanism of the magnetic tubes buoyancy (e.g.,
Fig.4a
\cite{ref41-3}), emerging as a result of the shear flows instability development in the tachocline, should be supplemented with the following results of the
Biermann postulate \cite{ref42-3} and the theory developed by Parker \cite{ref41-3,ref43-3}: the electric conductivity in the strongly ionized plasma may be so
high that the magnetic field becomes frozen into plasma and causes the split magnetic tube (Fig.~\ref{fig04-3}c,d) to cool inside.
%}}

\begin{figure}
\noindent
\begin{center}
\includegraphics[width=12cm]{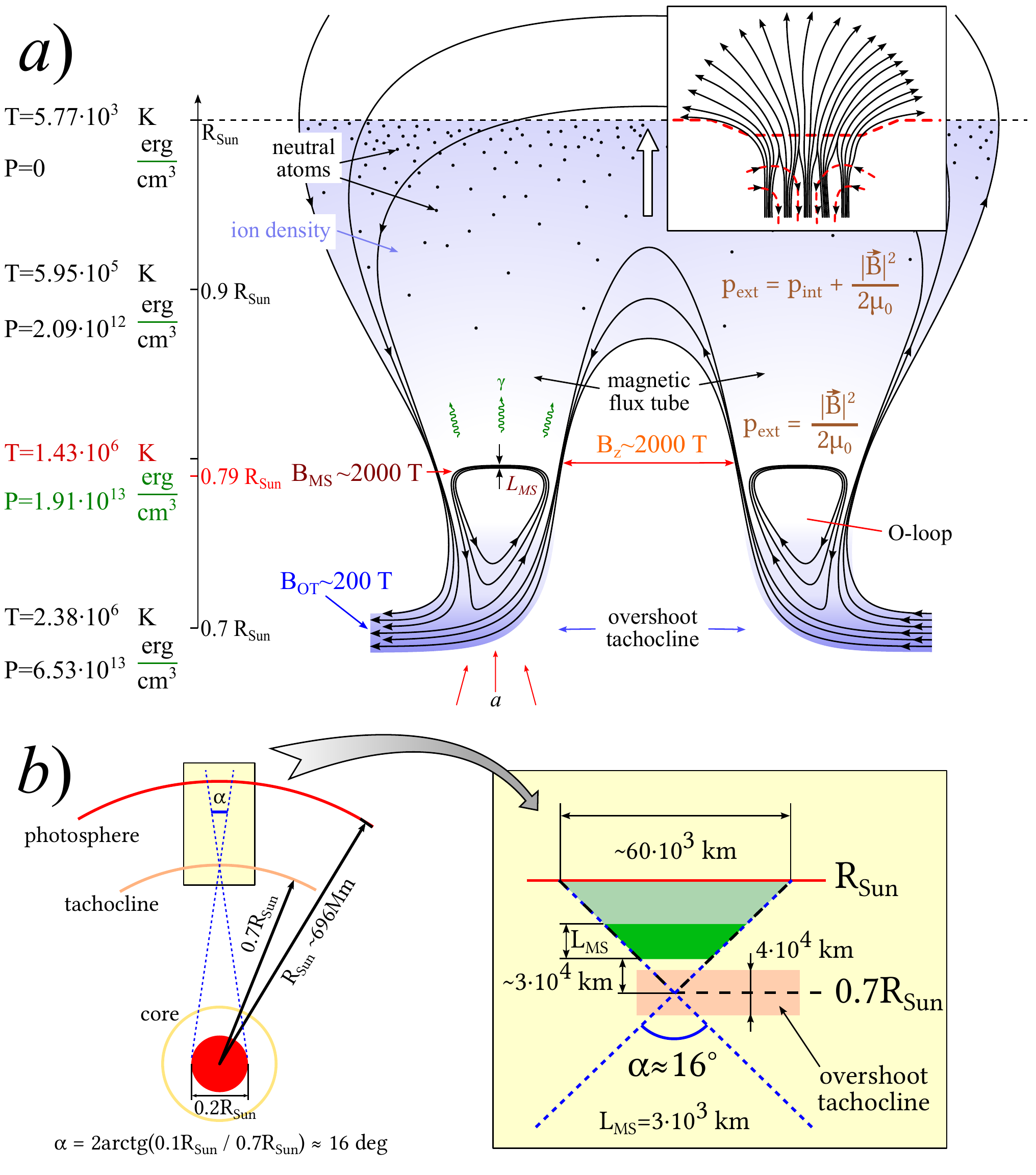}
\end{center}
\caption{(a)
% Magnetic flux tubes formation (black lines) in the tachocline through the
% shear instability development (more detailed than Fig.~\ref{fig04-3}d).
%The "capillary" effect in the magnetic tubes and a scheme of the axion (red
%arrows) conversion into $\gamma$-quanta (green arrows)
Topological effects of the magnetic reconnection inside the magnetic tubes with the "magnetic steps". The left panel shows the temperature and pressure change
along the radius of the Sun from the tachocline to the photosphere \cite{ref44-3,ref45-3}, $L_{MS}$ is the height of the magnetic shear steps (see
Fig.~\ref{fig04}c). Inset: a sketch of the magnetic field configuration, in which the field divides into individual flux tubes (or
"bundles"~\cite{Parker1994}) at some distance below the visible surface. The dashed arrows represent the presumed convective downdraft which helps to
hold the separate bundles together in the tight cluster that constitutes the sunspot~\cite{ref43-3}. \newline (b) All the axion flux, born via the
Primakoff effect (i.e. the real thermal photons interaction with the Coulomb field of the solar plasma) comes from the region  $\leq 0.1 R_{Sun}$~\cite{ref36}.
Using the angle $\alpha = 2 \arctan \left( 0.1 R_{Sun} / 0.7 R_{Sun} \right)$ marking the angular size of this region relative to tachocline, it is possible to
estimate the flux of the axions distributed over the surface of the Sun. The flux of the $\gamma$-quanta (of axion origin) is defined by the angle $\gamma = 2
\arctan \left( 0.5 d_{spot} / 0.3 R_{Sun} \right)$, where $d_{spot}$ is the diameter of a sunspot on the surface of the Sun (e.g., $d_{spot} \sim 60 \cdot 10^3
km$~\cite{Dikpati2008}).} \label{fig-lampochka}
\end{figure}

%\noindent \colorbox{yellow}{\parbox{\linewidth}{
Second, let us mention the known fact that the first neutral atoms appear in the convection zone at about $\sim$0.85$\,R_{Sun}$ -- helium at first, followed by
hydrogen. Their concentration grows towards the photosphere. It means that the upper boundary of the cool region in the split magnetic tube (e.g.,
Fig.~\ref{fig04-3}c) cannot exceed $0.85\,R_{Sun}$ (see e.g. Fig.~\ref{fig-lampochka}a). Otherwise the neutral atoms would penetrate freely along the
magnetic field lines towards the tachocline thus changing the heat transfer and the sunspot darkness \cite{ref34-3,ref35-3,ref36-3,ref37-3,ref38-3,ref39-3}.
%}}

%\noindent{\colorbox{yellow}{\parbox{\textwidth}{
%\begin{mdframed}[hidealllines=true,
%                 backgroundcolor=yellow,
%                 innerleftmargin=3pt,innerrightmargin=3pt,
%                 leftmargin=-3pt,rightmargin=-3pt]
In this regard let us consider, following~\cite{ref41-3}, the magnetic flux tube running horizontally through a gaseous electrically conducting medium
such as one finds in the Sun. It is well known that the tensile stress in the tube is $B^2 / (2 \mu)$ in mks units, where $B$ is the magnetic field density.
The magnetic field also exerts the outward pressure, and were the tube not impeded by the surrounding matter, it would expand. As it is, $B$ satisfies the
diffusion equation $\partial B / \partial t = \nu_m \nabla ^2 B$ where $\nu_m$ is the magnetic viscosity. If the medium is a sufficiently good conductor,
$\nu_m$ becomes small enough, so $\partial B / \partial t \cong 0$ and the field does not diffuse through the medium. Hydrostatic equilibrium requires that the
magnetic pressure $p_m$ is balanced by the gas pressure $p_{ext}$ outside the tube. Thus, if $p_{int}$ is the gas pressure inside the tube, we obtain the
desired relation $p_{ext} = p_{int} + p_m = p_{int} + B^2/ (2 \mu_0)$ (see (\ref{eq21})).
%\end{mdframed}
%}}

%\noindent \colorbox{yellow}{\parbox{\linewidth}{
In order to examine this question let us write down the ideal gas law

\begin{equation}
p = \dfrac{\rho k T}{m}
\label{eq25-3}
\end{equation}

\noindent and Eq. (\ref{eq21}) in terms of temperature and pressure. The result
may be represented in the following form  \cite{ref41-3}:

\begin{equation}
\rho_{int} - \rho_{ext} = \rho_{ext}\dfrac{ T_{ext} - T_{int}}{T_{int}} - \dfrac{m}{k T_{int}} \dfrac{B^2}{2 \mu},
\label{eq26-3}
\end{equation}
%}}

%\noindent \colorbox{yellow}{\parbox{\linewidth}{
\noindent where $\rho_{ext}$ and $\rho_{int}$ are the gas densities outside and inside the tube respectively, $k$ is the Boltzmann constant, and $m$ is the
mass of a single gas molecule.
%}}

%\noindent \colorbox{yellow}{\parbox{\linewidth}{
%Since the equation (\ref{eq26-3}) reaches the conditions of thermal equilibrium
%(\ref{eq21}) or disequilibrium (\ref{eq37}) under $T_{int} = T_{ext}$ and
%$T_{int} \rightarrow 0$ respectively, one may suppose (see Fig.~\ref{fig05-3}a)
%that the convective flows are suppressed by the magnetic tension inside the
%cool region (the split magnetic tube) deep in the convection zone, where the
%condition (\ref{eq37}) holds true. It means that the convective heat transfer is
%suppressed by the magnetic field  \cite{ref42-3}. On the contrary, in the
%near-surface region above $0.9 R_{Sun}$ the condition (\ref{eq21}) of the thermal
%equilibrium restores the "classic" energy transfer through convection.

Since Eq. (\ref{eq26-3}) reaches the condition of thermal disequilibrium (\ref{eq37}) with the surroundings when $T_{ext} \gg T_{int}$ or $T_{int} \rightarrow
0$, one may suppose (see Fig.~\ref{fig-lampochka}a) that the convective flows are suppressed by the magnetic tension inside the cool region (the split magnetic
tube) deep in the convection zone, where the condition (\ref{eq37}) holds true. It means that the convective heat transfer is suppressed by the magnetic
field~\cite{ref42-3}. According to~\cite{ref41-3}, let us assume that, applying the Parker-Biermann effect (or the cooling effect $(T_{ext} - T_{int})$), which
is simply proportional to the magnetic stresses with some constant factor $\kappa$, the value of $B^2 / (2\mu)$ may be estimated as follows:
%}}

%\noindent \colorbox{yellow}{\parbox{\linewidth}{
\begin{equation}
\frac{B^2 (z)}{2 \mu} = \frac{B^2 (0)}{2 \mu} \exp { \left[ \int \limits _0 ^z dz \frac{mg}{k T_{int}} \left( \frac{k \rho_{ext} \kappa}{m} - 1 \right) \right] },
\label{eq26-4}
\end{equation}
%}}

\noindent where $((k \rho_{ext} \kappa / m)-1)$ is integrated over distances several times the scale height $kT_{int} / (mg)$ and appears in the
exponent~\cite{ref41-3}. Thus, the fields of the order of $B(z) \sim 2000~T$ (see Fig.~\ref{fig-lampochka}a) are easily obtained from a field of $B(0) \sim
200~T$ (see Fig.~\ref{fig03} and Fig.~\ref{fig-lampochka}a) at the base of the flux tube, even though $((k \rho_{ext} \kappa / m)-1)$ may be only slightly
greater than zero.

\begin{figure}
\noindent
\begin{center}
\includegraphics[width=9cm]{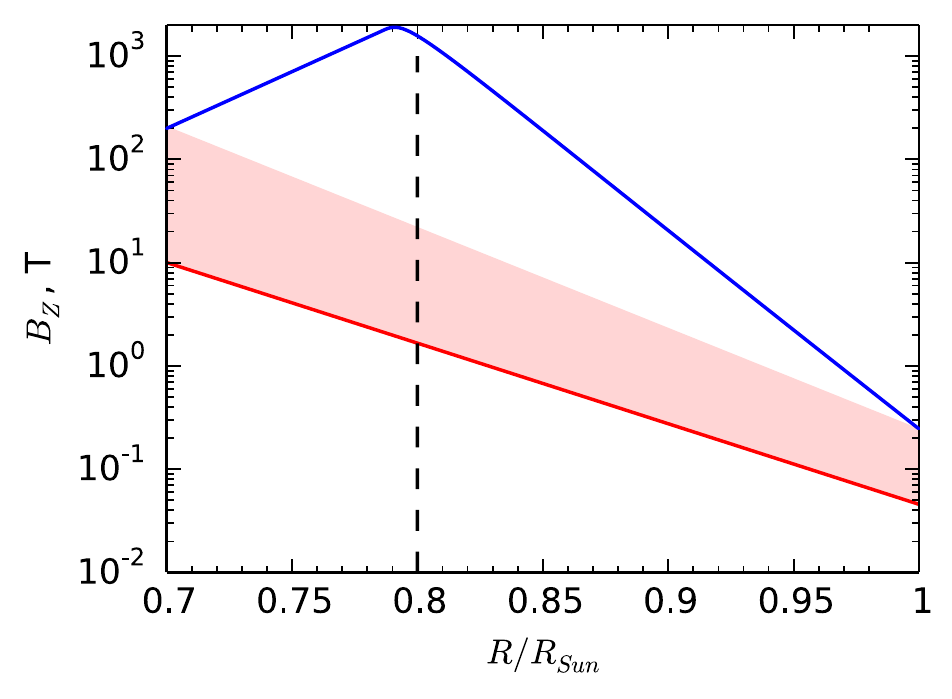}
\end{center}
\caption{Variation of the magnetic field strength $B_z$  along the emerging $\Omega$-loop as a function of the solar depth $R / R_{Sun}$  throughout the
convection zone. The blue and red lines mark the permitted values for (\ref{eq26-4})~\cite{ref41-3} and (\ref{eq26-5})~\cite{Parker1994} in the
 $B_z$~--~$R/R_{Sun}$ plane.}
\label{fig-Bz}
\end{figure}

One may also strengthen the cooling effect  $(T_{ext} - T_{int})$ dependence on the magnetic stress by introducing the square law of $\kappa (z)$. This yields
a solution (\ref{eq26-4}), which explains the behavior of the magnetic tension around the cool area inside the magnetic tube (see the blue curves on
Fig.~\ref{fig-Bz}).

%\noindent \colorbox{yellow}{\parbox{\linewidth}{
%\begin{mdframed}[hidealllines=true,
%                 backgroundcolor=yellow,
%                 innerleftmargin=3pt,innerrightmargin=3pt,
%                 leftmargin=0pt,rightmargin=-3pt]
On the contrary, in the convection zone above $0.85 R_{Sun}$ (see 
Fig.~\ref{fig-Bz}) the condition (\ref{eq21}) of the thermal 
non-equilibrium ($T_{ext} \neq T_{int}$) restores the "classic Parker" energy 
transfer to the surface through convection. As it was shown 
in~\cite{Parker1994} for this case, "...the observed 500~G\footnote{More 
up-to-date observations show the mean magnetic field strength at the level
$\sim$2500~G~\cite{Pevtsov2011,Pevtsov2014}.} intensity of the magnetic field 
emerging through the surface of the Sun can be understood from Bernoulli effect
in the upwelling $\Omega$-loop of magnetic field." From the general Parker 
ideology it also follows that the $B (0) > 10^4$~G azimuthal flux bundles above
$0.85 R_{Sun}$ (see Fig.~\ref{fig-Bz}) can be understood as a consequence of 
the large-scale buoyancy associated with the upwelling fluid in and around the
rising $\Omega$-loop, which fits in naturally with the Babcock-Leighton form of
the solar $\alpha\omega$-dynamo~\cite{Parker1994}.
As a result, the ideal $\Omega$-pumping process (above the cool region inside the tube) would eventually bring the field intensity of some of the flux bundles
by the following equation~\cite{Parker1994}:
%}}
%\noindent \colorbox{yellow}{\parbox{\linewidth}{
\begin{equation}
\frac{B^2 (z)}{8 \pi} = \left( \frac{B^2 (0)}{8 \pi} - p(0) I(z) \right) \exp {\left[ - \int \limits _0 ^z \frac{dz}{\Lambda (z)} \right]},
\label{eq26-5}
\end{equation}
%}}

\noindent where $\Lambda (z) = k T(z) / (\mu M g)$ is the pressure scale height, $\mu$ is the mean molecular weight, $M$ is the mass of the hydrogen atom and
$I$ is the dimensionless integral

%\noindent \colorbox{yellow}{\parbox{\linewidth}{
\begin{equation}
I (z) = \int \limits _0 ^z \frac{dz}{\Lambda (z)} \frac{\Delta T (z)}{T(z)}
\end{equation}
%}}

\noindent representing the extra buoyancy of the elevated temperature $\Delta T(z)$ of the upward billowing gas within and around the magnetic flux
bundle~\cite{Parker1994}.

It is noteworthy that the qualitative nature of the $\Omega$-loop formation and growth process, based on the semiphenomenological model of the magnetic
$\Omega$-loops in the convective zone, may be seen as follows.

\vspace{0.3cm}

%\noindent \colorbox{yellow}{\parbox{\linewidth}{
%\begin{itemize}
%\item
\noindent $\bullet$ A high concentration azimuthal magnetic flux ($B_{OT} \sim 200$~T, see Fig.~\ref{fig03} and Fig.~\ref{fig-Bz}) in the overshoot tachocline
through the shear flows instability development.
%}}

An interpretation of such link is related to the fact that helioseismology places the principal rotation $\partial \omega / \partial r$ of the Sun in the
overshoot layer immediately below the bottom of the convective zone~\cite{Parker1994}. It is also generally believed that the azimuthal magnetic field of the
Sun is produced by the shearing $r \partial \omega / \partial r$ of the poloidal field $B_p$ from which it is generally concluded that the principal azimuthal
magnetic flux resides in the shear layer~\cite{ref16,Parker1993}.

\vspace{0.3cm}

\noindent
%\item
$\bullet$ If some "external" factor of the local (magnetic?) shear perturbation appears against the background of the azimuthal magnetic flux concentration,
such additional local density of the magnetic flux may lead to the magnetic field strength as high as, e.g., $B_z \sim 2000$~T (see Fig.~\ref{fig-lampochka}a).
Of course, this brings up a question about the physics behind such "external" factor and the local shear perturbation.

%\noindent \colorbox{yellow}{\parbox{\linewidth}{
In this regard let us consider the superintense magnetic $\Omega$-loop formation in the overshoot tachocline through the local shear caused by the high local
concentration of the azimuthal magnetic flux. The buoyant force acting on the $\Omega$-loop grows with this concentration so the vertical magnetic field of the
$\Omega$-loop reaches $B_z \sim 2000$~T at about $R / R_{Sun} \sim 0.79$. Because of the magnetic pressure (\ref{eq37}) $p_{ext} = 1.91\cdot 10^{13}
~erg/cm^3$~\cite{ref44-3,ref45-3} this leads to a significant cooling of the $\Omega$-loop tube.
%}}

In other words, we assume the effect of the $\Omega$-loop cooling to be the basic effect responsible for the magnetic flux concentration. It arises from the
well known suppression of convective heat transport by a strong magnetic field~\cite{ref42-3}. It means that although the principal azimuthal magnetic flux
resides in the shear layer, it predetermines the additional local shear giving rise to a significant cooling inside the $\Omega$-loop.

%\noindent \colorbox{yellow}{\parbox{\linewidth}{
Thus, the ultralow pressure is set inside the magnetic tube as a result of the sharp limitation of the magnetic steps buoyancy inside the cool magnetic tube
(Fig.~\ref{fig-lampochka}a). This happens because the buoyancy of the magnetic flows requires finite superadiabaticity of the convection
zone~\cite{ref47-3,ref35-3}, otherwise, expanding according to the magnetic adiabatic law (with the convection being suppressed by the magnetic field), the
magnetic clusters may become cooler than their surroundings, which compensates the effect of the magnetic buoyancy.
%}}

%\noindent \colorbox{yellow}{\parbox{\linewidth}{
Eventually we suppose that the axion mechanism based on the $\gamma$-quanta channeling along the "cool" region of the split magnetic tube
(Fig.~\ref{fig-lampochka}a) effectively supplies the necessary heat flux channeling to the photosphere while the convective heat transfer is heavily
suppressed.
%}}
%\noindent \colorbox{yellow}{\parbox{\linewidth}{
%\colorbox{red}{to a Footnote?}
From here it follows that all physics of the "local shear -- strong cooling -- axion mechanism of $\gamma$-quanta channeling" link is based on the local shear
variations, which is unfortunately currently unexplored.\footnote{Our idea is that the local azimuthal magnetic flux perturbations may arise in the overshoot
tachocline of the Sun as a result of the interaction between the nuclei (mainly protons) within the Sun and asymmetric dark matter~\cite{Frandsen2010}.
The variations of the dark matter flow intensity may be driven by the stars orbiting around the massive black hole in the center of the Milky Way (see,
e.g.~\cite{Gillessen2009}). However, that is a subject of our forthcoming paper.}
%}

\vspace{0.3cm}

\noindent
%\colorbox{yellow}{\parbox{\linewidth}{
%\item
$\bullet$ Under the significant growth of the inner gas pressure above $0.79 \div 0.85 R/R_{Sun}$ (see Fig.~\ref{fig-lampochka} and Fig.~\ref{fig-Bz}),
according to the Parker's ideology of the azimuthal magnetic flux bundles~\cite{Parker1994}, rather intense bundles appear inside the $\Omega$-loop as a
consequence of the large-scale buoyancy associated with the upwelling fluid in and around the rising $\Omega$-loop (see the inset in
Fig.~\ref{fig-lampochka}a).
%}}

\vspace{0.3cm}

\noindent
%\colorbox{yellow}{\parbox{\linewidth}{
%\item
$\bullet$ It is not difficult to show that the convective updraft set in motion by the buoyant rise of the magnetic $\Omega$-loop drives an upward convective
flow extending all the way across the upper part of the convective zone, i.e. from the level above the cool region inside the magnetic tube (see the blue line
in Fig.~\ref{fig-Bz}) to the visible surface, sketched in Fig.~\ref{fig-lampochka}.
%}}

\vspace{0.3cm}

\noindent
%\colorbox{yellow}{\parbox{\linewidth}{
%\item
$\bullet$ The vigor of the $\Omega$-loops increases with the increasing concentration of the azimuthal flux bundles, so that the azimuthal flux bundles, formed
above the cool region inside the tube, are eventually concentrated to $B_z > 10^4 ~G$ (see the blue line in Fig.~\ref{fig-Bz}) or more at low latitudes, where
only the bundles of the intensity are able to rise the $\Omega$-loop in opposition to the stabilizing effect of the strong shear and subadiabatic temperature
gradient~\cite{Parker1994,Spruit1974}.
%}}

\vspace{0.3cm}

\noindent
%\colorbox{yellow}{\parbox{\linewidth}{
%\item
$\bullet$ As a consequence, the upward displacement of the plasma within the $\Omega$-loop in the superadiabatic temperature gradient of the upper part of the
convective zone causes the plasma to push upward in the vertical legs of the $\Omega$-loop, thereby pulling plasma out of the horizontal azimuthal flux, which
although appears above the cool region inside the magnetic tube, is indirectly connected with each foot of the $\Omega$-loop resting in the overshoot
tachocline. This introduces an ad-hoc constraint on the footpoint of the flux loops~\cite{DSilva1993}.
%\end{itemize}
%}}

\vspace{0.3cm}

And finally, it may be supposed that the electric conductivity in the heavily ionized plasma is indeed very high, and the magnetic field is "frozen into"
plasma, strictly suppressing the convective heat transfer. At the same time, the condition (\ref{eq26-3}) of the thermal non-equilibrium ($T_{int} \neq
T_{ext}$) reestablishes the convection above $0.85 R_{Sun}$ up to the visible surface, preventing the neutral atoms penetration into the deeper layers along
the magnetic tubes (see Fig.~\ref{fig-lampochka}).

Some kind of topological effects of the magnetic reconnection inside the magnetic tubes will be observed in this case (see Fig.~\ref{fig04}b and
Fig.~\ref{fig-lampochka}a). A complete representation of the spatio-temporal evolution of the magnetic tube in the convection zone presented in
Fig.~\ref{fig04-3}a illustrates the process of axions conversion into $\gamma$-quanta (within the height $L_{MS}$ of the magnetic "steps") and the active
regions formation in the solar photosphere (Fig.~\ref{fig04}d), thus shedding light on the unique mechanism of energy transport along the magnetic tubes
between the overshoot tachocline and the photosphere.

%\noindent \colorbox{yellow}{\parbox{\linewidth}{
We claim that the axion mechanism of Sun luminosity is closely related to the structure of a sunspot \cite{ref49-3}, inside of which the magnetic field
produces a "cool" region of the split magnetic tube (e.g., Fig.~\ref{fig04-3}d), thus inhibiting the convective heat transfer.
%}}

%\noindent \colorbox{yellow}{\parbox{\linewidth}{
As it is known, it was Deinzer \cite{ref51-3} who drew attention to the fact that the convection could not be suppressed completely, since the temperature of
the sunspot umbrae was as high as 4000 K, which could not be explained by the radiative transfer and heat conduction only. It is also related to the Jahn's
idea \cite{ref52-3}, which explains the coolness of the spot through the heat flux channeling along the magnetic field lines, as it was suggested by Hoyle
\cite{ref12}. The global models of the sunspots basing on this concept were studied by Chirte \cite{ref13} and Chirte \& Shaviv \cite{ref55-3}, who concluded
that the convection inhibition (suggested by Biermann \cite{ref42-3}) and the heat flux channeling \cite{ref12} indeed may take place in the sunspots.
%}}

%%\noindent \colorbox{yellow}{\parbox{\linewidth}{
%Developing these ideas further, we suppose that the axion mechanism based on the $\gamma$-quanta channeling along the "cool" region of the split magnetic tube
%(Fig.~\ref{fig-lampochka}a) effectively supplies the necessary heat flux to the photosphere while the convective heat transfer is heavily suppressed. Taking into
%account the fact that the magnetic tube is in thermal equilibrium~(\ref{eq21}) with its surroundings above $0.9 R_{Sun}$, according to Fig.~\ref{fig-lampochka}a,
%the
%energy transfer occurs through the overturning convection between the $0.9 R_{Sun}$ and the surface.
%%}}

Developing these ideas further, we suppose that the axion mechanism based on the $\gamma$-quanta channeling along the "cool" region of the split magnetic tube
(Fig.~\ref{fig-lampochka}a) effectively supplies the necessary heat flux to the photosphere while the convective heat transfer is heavily suppressed. On the
other hand, the rising magnetic tube above $0.85 R_{Sun}$ is under conditions (\ref{eq26-3}) of thermal non-equilibrium ($T_{ext} \neq T_{int}$). It means that
starting at $0.85 R_{Sun}$ and up to the surface of the photosphere the energy transfer is provided not only by the heat flux channeling, but also by the
reestablished convection. As a result, it may be concluded that the axion mechanism of Sun luminosity based on the lossless $\gamma$-quanta channeling along
the magnetic tubes allows to explain the effect of the partial suppression of the convective heat transfer, and thus to understand the known puzzling darkness
of the sunspots~\cite{ref34-3}.

%\subsection{Estimation of the axion-photon oscillation parameters}
\subsection{Estimation of the solar axion-photon oscillation parameters on the 
            basis of the hadron axion-photon coupling in white dwarf cooling}
\label{subsec-osc-parameters}

%Let us consider the modulation of an axion flux emerging from the Sun core on passing through the solar tachocline region (ST) located beneath the base of the
%solar convective zone (Fig.~\ref{fig04}b). It is known that the thickness $\Delta_{ST}$ of ST obtained by Charbonneau~\textit{et al.}~\cite{ref44} is
%$\Delta_{ST} / R_S = 0.039 \pm 0.013$ at the equator and $\Delta_{ST} / R_S = 0.042 \pm 0.013$ at a latitude of 60$^{\circ}$, suggesting that the tachocline
%may get somewhat wider at high latitudes but that the result is not statistically significant.
%
%Taking into account that the value of the magnetic field strength $B_{OT}$ in the overshoot tachocline is $\sim 400 ~T$ (see Fig.~\ref{fig03}) and setting the
%overshoot tachocline thickness $L_{OT}$ approximately equal $\sim 3 \cdot 10^4 ~km$ (see Fig.~\ref{fig04}b) it is not hard to use the expression (\ref{eq13})
%in the form
%\begin{equation}
%\left( \frac{9 \cdot 10^{10} ~km}{L_{OT}} \right) \left( \frac{1 G}{B_{OT}} \right) \left( \frac{10^{-10} GeV^{-1}}{g_{a\gamma}} \right) \sim 1
%\label{eq38}
%\end{equation}
%for estimating the axion coupling constant to photons
%\begin{equation}
%g_{a \gamma} \sim 7.2 \cdot 10^{-11} ~~ GeV^{-1}\, . \label{eq39}
%\end{equation}

%\noindent \colorbox{yellow}{\parbox{\linewidth}{
It is known~\cite{Cadamuro2012} that astrophysics provides a very interesting clue concerning the evolution of white dwarf stars with their small mass
predetermined by the relatively simple cooling process. It is related to the fact that recently it has been possible to determine their luminosity function
with the unprecedented precision~\cite{Isern2008}. It seems that if the axion has a direct coupling to electrons and a decay constant $f_a \sim 10^{9} ~GeV$,
it provides an additional energy-loss channel that permits to obtain a cooling rate that better fits the white dwarf luminosity function than the standard
one~\cite{Isern2008}. The hadronic axion (with the mass in the $meV$ range and $g_{a\gamma \gamma} \sim 10^{-12} ~GeV^{-1}$) would also help in fitting the
data, but in this case a stronger value for $g_{a\gamma \gamma}$ is required to perturbatively produce an electron coupling of the required strength.
%}}

\begin{figure*}
%\centerline{\includegraphics[width=9cm]{gagamma_ma-limits.eps}}
\centerline{\includegraphics[width=9cm]{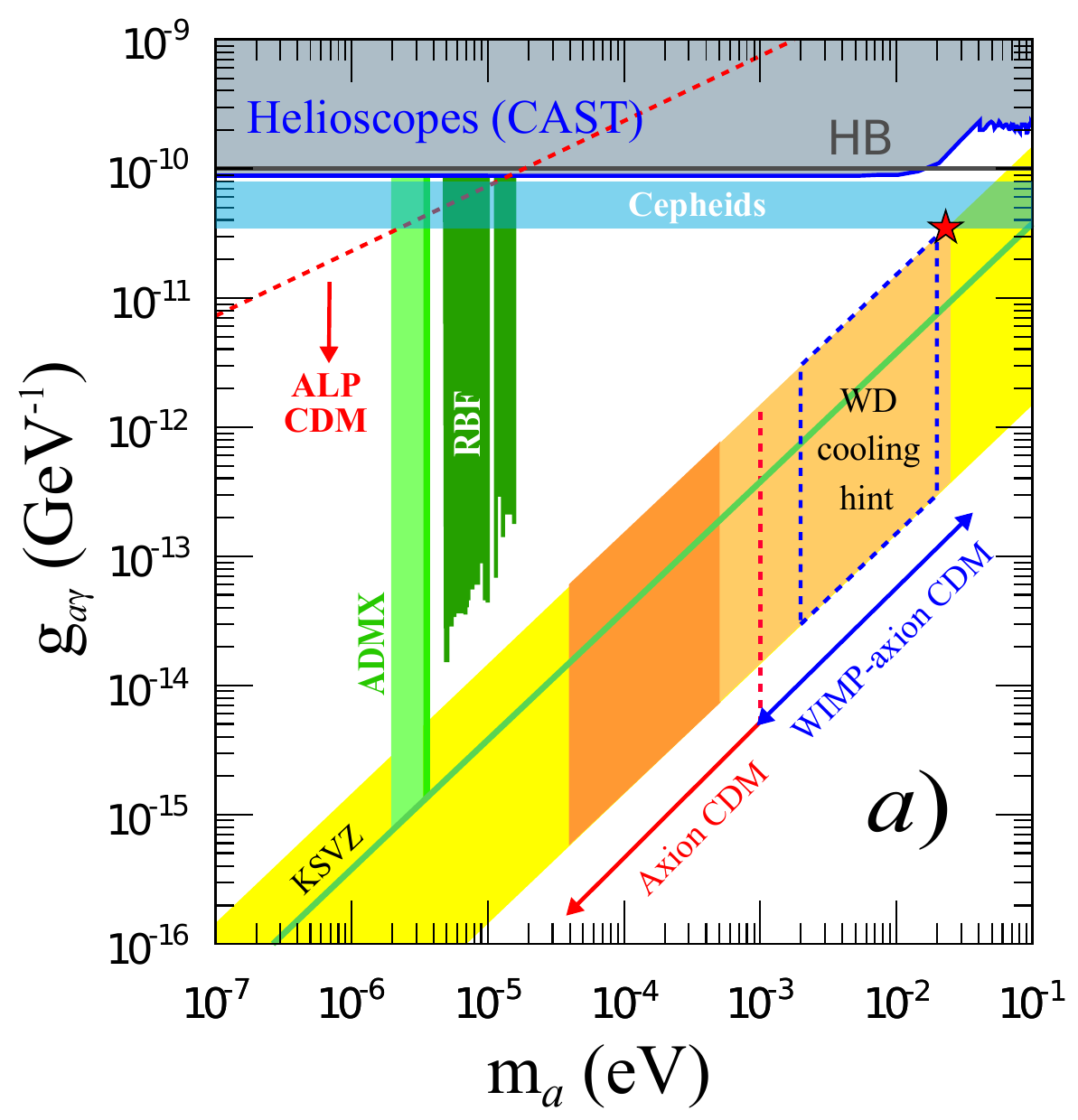}}

\caption{Summary of astrophysical, cosmological and laboratory constraints on 
axions and axion-like particles. Comprehensive axion/ALP parameter space,
highlighting the two main front lines of direct detection experiments: 
helioscopes (CAST~\cite{ref58,ref72,CAST2011,Arik2013}) and haloscopes 
(ADMX~\cite{ref50} and RBF~\cite{ref51}). The astrophysical bounds from 
horizontal branch and massive stars are labeled "HB"~\cite{ref02} and 
"Cepheids"~\cite{Carosi2013} respectively. The QCD motivated models 
(KSVZ~\cite{ref46,ref46a} and DFSZ~\cite{ref47}) for axions lay in the yellow 
diagonal band. The orange parts of the band correspond to cosmologically 
interesting axion models: models in the "classical axion window" possibly 
composing the totality of DM (labelled "Axion CDM") or a fraction of it 
("WIMP-axion CDM"~\cite{Baer2011}). For more generic ALPs, practically all the
allowed space up to the red dash line may contain valid ALP CDM 
models~\cite{Arias2012}. The region of axion masses invoked in the WD cooling 
anomaly is shown by the blue dash line~\cite{Irastorza2013}.
The red star marks the values of the axion mass $m_a \sim 2.3 \cdot 10^{-2}
eV$ and the axion-photon coupling constant $g_{a\gamma} \sim 3.6 \cdot 10^{-11}
GeV^{-1}$ chosen in the present paper on the basis of the suggested relation
between the axion mechanisms of the Sun's and the white dwarf luminosity.}
\label{fig05}
\end{figure*}

%\noindent \colorbox{yellow}{\parbox{\linewidth}{
Our aim is to estimate the solar axion-photon oscillation parameters basing on the hadron axion-photon coupling derived from white dwarf cooling. The estimate
of the horizontal magnetic field in the O-loop is not related to the photon-axion conversion in the Sun only, but also to the axions in the model of white
dwarf evolution. Therefore along with the values of the magnetic field strength $B_{MS} \sim 2 \cdot 10^3 ~T$ and the height of the magnetic shear steps
$L_{MS} \sim 3 \cdot 10^3 ~km$ (Fig.~\ref{fig-lampochka}a,b) we use the following parameters of the hadronic axion (from the White Dwarf area in
Fig.~\ref{fig05}a~\cite{Irastorza2013, Carosi2013}):
%}}

%\noindent \colorbox{yellow}{\parbox{\linewidth}{
\begin{equation}
g_{a \gamma} \sim 3.6 \cdot 10^{-11} ~ GeV^{-1}, ~~~ m_a \sim 2.3 \cdot 10^{-2} ~eV.
\label{eq3.30}
\end{equation}
%}}

%\noindent \colorbox{yellow}{\parbox{\linewidth}{
The choice of these values is also related to the observed solar luminosity variations in the X-ray band (see (\ref{eq3.35})). The theoretical consequences of
such choice are considered below.
%}}

%\noindent \colorbox{yellow}{\parbox{\linewidth}{
Let us remind that the magnetic field strength $B_{OT}$ in the overshoot tachocline of $\sim 200~T$ (see Fig.~\ref{fig03}) and the Parker-Biermann cooling
effect ($T_{ext} - T_{int}$) in the form of the square law $\kappa (z)$ in (\ref{eq26-4}) lead to the corresponding value of the magnetic field strength $B(z =
0.79 R_{Sun}) \sim 2000 ~T$ (see Fig.~\ref{fig-lampochka}a and Fig.~\ref{fig-Bz}), which in its turn assumes virtually zero internal gas pressure of the
magnetic tube.
%}}

%\noindent \colorbox{yellow}{\parbox{\linewidth}{
As it is shown above (see~\ref{subsec-ideal-channeling}), the topological effect of the magnetic reconnection
inside the $\Omega$-loop results in the formation of the so-called O-loops (Fig.~\ref{fig04}c and Fig.~\ref{fig-lampochka}a) with their buoyancy limited from
above by the strong cooling inside the $\Omega$-loop (Fig.~\ref{fig-lampochka}a). It is possible to derive the value of the horizontal magnetic field of the
magnetic steps at the top of the O-loop ($L_{MS} \sim 3 \cdot 10^3 ~km$): $\vert B_{MS} \vert \approx \vert B(z = 0.79 R_{Sun}) \vert \sim 2000 ~T$.
%}}

%\noindent \colorbox{yellow}{\parbox{\linewidth}{
Basing on the assumptions associated with the measured white dwarf luminosity function~\cite{Isern2008}, it is not hard to use the expression
(\ref{eq08}) for the conversion probability\footnote{Hereinafter we use rationalized natural units to convert the magnetic field units from $Tesla$ to
$eV^2$, and the conversion reads $1\,T = 195\,eV^2$~\cite{Guendelman2009}.}

%\noindent \colorbox{yellow}{\parbox{\linewidth}{
\begin{equation}
P_{a \rightarrow \gamma} = \frac{1}{4} \left( g_{a \gamma} B_{MS} L_{MS} \right)^2 \sim 0.0114
\label{eq3.31}
\end{equation}
%}}
%\noindent \colorbox{yellow}{\parbox{\linewidth}{
for estimating the axion coupling constant to photons (\ref{eq3.30}).
%}}

%q = \frac{\left \vert m_{\gamma}^2 - m_{a}^2 \right \vert}{2E_a}
%

It is necessary to make a digression concerning some important features of the oscillation length ($L_{osc} = 2 \pi / q$) here. It is known that in order to
maintain the maximum conversion probability, i.e. zero momentum transfer ($q =\left \vert m_{\gamma}^2 - m_{a}^2 \right \vert / (2E_a);\ q \to 0$), the axion
and photon fields, put into some medium ($m_{\gamma} \equiv m_{a}$), need to remain in phase over the length of the magnetic field. This coherence condition is
met when $q L_{MS} \leqslant \pi$, and allows one to obtain the following remarkable relation~\cite{ref36} between the medium density variations and
axion mass variations for the coherent case $q \to 0$, i.e. $m_{\gamma} \equiv m_a$:

%\noindent \colorbox{yellow}{\parbox{\linewidth}{
\begin{equation}
\frac{\Delta \rho}{\rho} = 2 \frac{\Delta m_a}{m_a} = \frac{4 \pi E_a}{m_a^2 L_{MS}}\, . \label{eq3.31b}
\end{equation}
%}}

It is easy to show that for the mean energy $E_a = 4.2 ~keV$, the axion mass $m_a = 2.3 \cdot 10^{-2}~eV$ and the thickness of the magnetic steps $L_{MS} = 3
\cdot 10^6 ~m$ the density variations in~(\ref{eq3.31b}) are $\sim$10$^{-6}$. It means that the inverse coherent Primakoff effect takes place
only when the variations of the medium density inside a cylindric volume of the "height" $L_{MS}$ (see Fig.~\ref{fig04}b) do not exceed the value
$\sim$10$^{-6}$. Although this is a very strong restriction, the physical process in the magnetic flux tube (see Fig.~\ref{fig04}b) would "freeze" the
plasma (low-Z gas) in this magnetic volume so much that this restriction on the density variations is fulfilled.

%And finally let us give the estimates for the allowed axion masses which may be
%obtained on the basis of the linear limitations (\colorbox{red}{\ref{eq10}}) and (\ref{eq11})
%with account of (\ref{eq3.31}). The analysis of the upper linear limitations
%(Eq.~(\ref{eq10})) and the lower ones (Eq.~(\ref{eq11})) shows that the axion
%masses
%%$3.5 \cdot 10^{-6} ~eV$,  $10^{-5} ~eV$ and $1.7 \cdot 10^{-5} ~eV$ are
%\hl{$2.3 \cdot 10^{-2} ~eV$ is}
%in good agreement with these limits (see Fig.~\ref{fig05}b).On the other hand,
%these mass values are obviously in a good agreement with the known experimental
%limitations shown in Fig.~\ref{fig05}a in the $m_a$ -- $g_{a\gamma}$ plane.

Thus, it is shown that the hypothesis about the possibility for the solar axions born in the core of the Sun to be efficiently converted back into
$\gamma$-quanta in the magnetic field of the
%solar overshoot tachocline
magnetic steps of the O-loop (above the solar overshoot tachocline) is relevant. Here the variations of the magnetic field in the solar tachocline are the
direct cause of the converted $\gamma$-quanta intensity variations. The latter in their turn may be the cause of the overall solar luminosity variations known
as the active and quiet Sun phases.

\

%\section{Axion mechanism of TSI variations}
%\label{sec-tsi-variations}
%
%
%It is known~\cite{ref53,ref54} that during the transition from the minimum to the maximum of the 11-year magnetic solar cycle, i.e. from the quiet to the
%active phase of the Sun, the relative TSI variations ($\Delta$TSI) are
%\begin{equation}
%\frac{\vert \Delta TSI \vert}{TSI} \sim 5 \cdot 10^{-4}\, . \label{eq40}
%\end{equation}
%
%As the theoretical calculations show~\cite{ref54}, such effect of solar activity modulation is associated with the magnetic energy release from the dynamo
%region to the photosphere and higher with the consequent transformation into radiation. According to~\cite{ref54}, the surface (over-photosphere) magnetic
%fields are considered as the additional channels for energy transport from the inner layers of the Sun outwards in this case. Interestingly, such mechanism of
%Sun modulation is corroborated by the results of the papers~\cite{ref55,ref56,ref57}, which conclude that the luminosity growth during the solar cycle maximum
%may be associated with the effective over-photosphere magnetic fields influence on the electromagnetic radiation output.

%On the other hand,

It is easy to show that the theoretical estimate for the part of the axion luminosity $L_a$ in the total luminosity of the Sun $L_{Sun}$ with respect to
(\ref{eq3.30}) is~\cite{ref58}
\begin{equation}
\frac{L_a}{L_{Sun}} = 1.85 \cdot 10 ^{-3} \left( \frac{g_{a \gamma}}{10^{-10} GeV^{-1}} \right)^2 \sim 2.5 \cdot 10^{-4} .
\label{eq3.32}
\end{equation}

\begin{figure*}
%\centerline{\includegraphics[width=12cm]{Sun_total_spectra-10.eps}}
\centerline{\includegraphics[width=12cm]{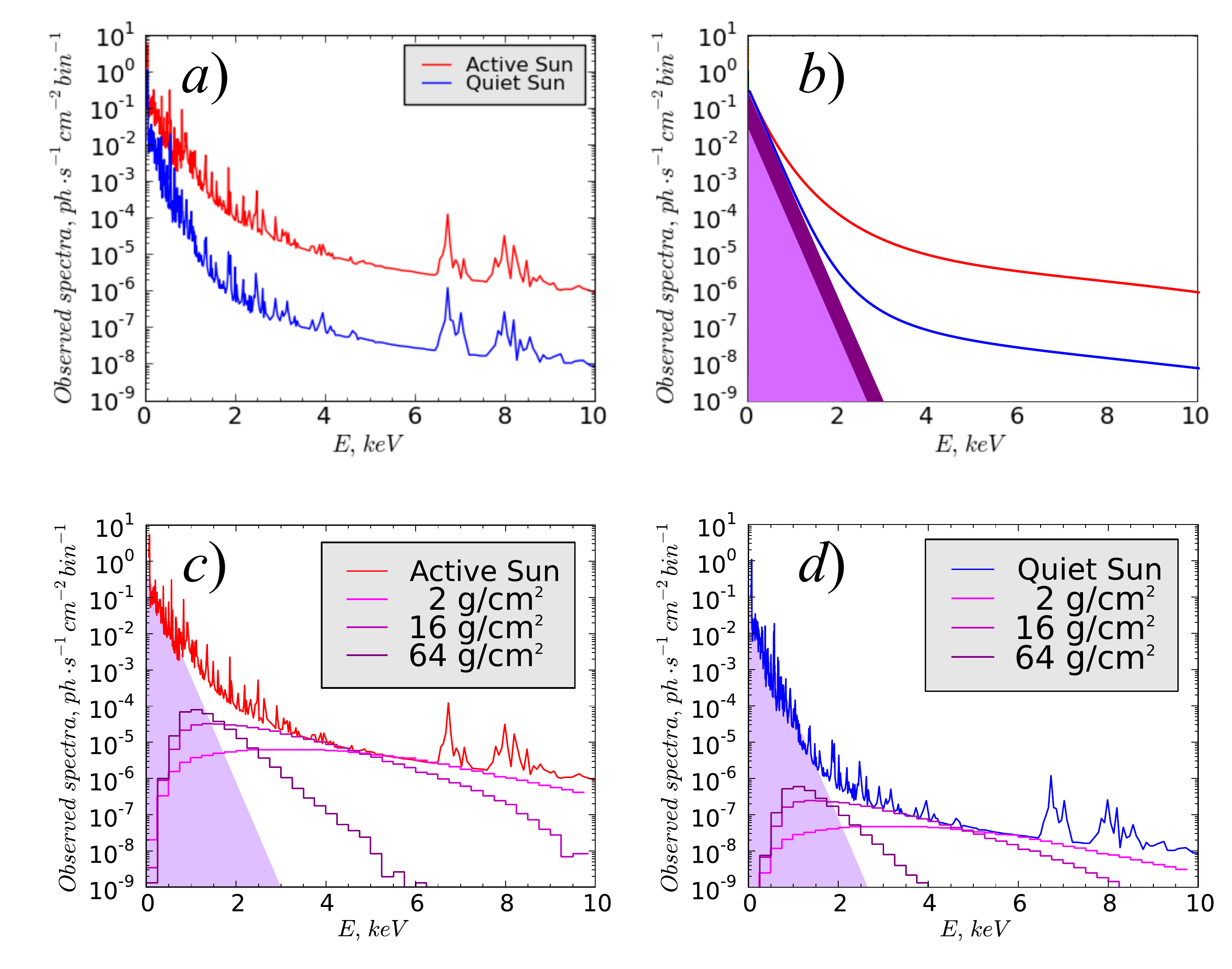}}

\caption{(a) Reconstructed solar photon spectrum below 10~keV from the active Sun (red line) and quiet Sun (blue line) from accumulated observations (spectral
bin is 6.1~eV wide). Adopted from~\cite{ref59}. \newline (b) Axion mechanism of luminosity variations: the soft part of the solar photon spectrum is invariant,
and thus does not depend on the phase of the Sun (violet band), while the red and blue curves characterize the luminosity increment in the active and quiet
phases. \newline (c) Reconstructed solar photon spectrum fit in the active phase of the Sun by the quasi-invariant soft part of the solar photon spectrum and
three spectra (\ref{eq3.33}) degraded to the Compton scattering for column densities above the initial conversion place of 64, 16 (adopted from~\cite{ref36})
and 2 $g / cm^2$ (present paper).
%The converted solar axion spectrum (CSAS) with the coefficient $1.5 \cdot 10^{-16}$ ( see explanation in the text) is also shown for comparison;
\newline (d) The similar curves for the quiet phase of the Sun. \label{fig06}}
\end{figure*}

%\noindent \colorbox{yellow}{\parbox{\linewidth}{
%\sout{It should be mentioned that the comparison of the estimates (\ref{eq40}) and (\ref{eq41}) shows that the portion of the axions that are converted into
%$\gamma$-quanta does not exceed 50\% which is highly important for the solar axion flux estimations performed in the real ground-based experiments.}}}
%At the same time, despite the coincidence of the relative variations (\ref{eq40}) and (\ref{eq41}) by the order of magnitude, one should also keep in mind
% that the axion mechanism of Sun luminosity variations is somewhat different by the nature of its source, although it is also associated with the radiant
%energy input to the photosphere.
As opposed to the classic mechanism of the Sun modulation mentioned above, the axion mechanism is determined by the magnetic tubes rising to
the photosphere, and not by the over-photosphere magnetic fields.  In this case the solar luminosity modulation is determined by the axion-photon oscillations
%in the magnetic field of the overshoot tachocline
in the magnetic steps of the O-loop causing the formation and channeling of the $\gamma$-quanta inside
%the almost empty magnetic tubes (see Fig.~\ref{fig04}b).
the almost empty magnetic $\Omega$-tubes (see Fig.~\ref{fig04} and Fig.~\ref{fig-lampochka}a). When the magnetic tubes cross the photosphere,
they "open" (Fig.~\ref{fig-lampochka}a), and the $\gamma$-quanta are ejected to the photosphere, where their comfortable journey along the magnetic
tubes (without absorption and scattering) ends. As the calculations by Zioutas~\textit{et al.}~\cite{ref36} show, the further destiny of the $\gamma$-quanta in
the photosphere may be described by the Compton scattering, which actually agrees with the observed solar spectral shape (Fig.~\ref{fig06}c,d).

From the axion mechanism point of view it means that the solar spectra during the active and quiet phases (i.e. during the maximum and minimum solar activity)
differ from each other by the smaller or larger part of the Compton spectrum, the latter being produced by the $\gamma$-quanta of the axion origin ejected
from the magnetic tubes into the photosphere.

A natural question arises at this point: "What are the real parts of the Compton spectrum of the axion origin in the active and quiet phases of the Sun
respectively, and do they agree with the experiment?" Let us perform the mentioned estimations basing on the known experimental results by ROSAT/PSPC, where
the Sun's coronal X-ray spectra and the total luminosity during the minimum and maximum of the solar coronal activity were obtained~\cite{ref59}.

%\noindent \colorbox{yellow}{\parbox{\linewidth}{
Apparently, the solar photon spectrum below 10~keV of the active and quiet Sun (Fig.~\ref{fig06}a) reconstructed from the accumulated ROSAT/PSPC observations
may be described by three Compton spectra for different column densities rather well (Fig.~\ref{fig06}c,d). This gives grounds for the assumption that the hard
part of the solar spectrum is mainly determined by the axion-photon conversion efficiency.
%}}

With this in mind, let us suppose that the part of the differential solar axion flux at the Earth~\cite{ref58}

%\noindent \colorbox{yellow}{\parbox{\linewidth}{
\begin{align}
\frac{d \Phi _a}{dE} = 6.02 \cdot 10^{10} \left( \frac{g_{a\gamma}}{10^{10} GeV^{-1}} \right)^2 E^{2.481} \exp \left( - \frac{E}{1.205} \right) ~~cm^{-2}
s^{-1} keV^{-1} ,
\label{eq3.33}
\end{align}
%}}
%\noindent \colorbox{yellow}{\parbox{\linewidth}{
%\noindent \colorbox{yellow}{\parbox{\linewidth}{
which characterizes the differential $\gamma$-spectrum of the axion origin $d \Phi _{\gamma} / dE$ (Fig.~\ref{fig06}c,d), i.e.
%}}
%}}
%\noindent \colorbox{yellow}{\parbox{\linewidth}{
%\begin{align}
%\frac{d \Phi _{\gamma}}{dE}  \cong \frac{1}{3} P_{a \leftrightarrow \gamma} \frac{d \Phi _{a}}{dE} ~~ photon \cdot cm^{-2} s^{-1} bin^{-1} \cong  2 \cdot
%10^{-3} \frac{d \Phi _{a}}{dE} ~~ ph \cdot cm^{-2} s^{-1} bin^{-1}\, , \label{eq43}
%\end{align}
%\noindent \colorbox{yellow}{\parbox{\linewidth}{
\begin{align}
\frac{d \Phi _{\gamma}}{dE}  \cong P_{\gamma} \frac{d \Phi _{a}}{dE} ~~ cm^{-2} s^{-1} keV^{-1} \approx 3 \cdot 10^{-5}
\frac{d \Phi _{a}}{dE} ~ ph\cdot cm^{-2} s^{-1} bin^{-1}
\label{eq3.34}
\end{align}
%}}
%}}
%\noindent \colorbox{yellow}{\parbox{\linewidth}{
where the spectral bin width is 6.1~eV (see Fig.~\ref{fig06}a); the probability $P_{\gamma}$
describing the relative portion of $\gamma$-quanta (of axion origin) channeling along the magnetic tubes may be defined, according to~\cite{ref59}, from the
observed solar luminosity variations in the X-ray band, recorded in ROSAT/PSPC experiments (Fig.~\ref{fig06}): $\left(L_{corona}^X \right) _{min} \approx 2.7
\cdot 10^{26} ~erg/s$ at minimum and $\left( L_{corona}^X \right) _{max} \approx 4.7 \cdot 10^{27} ~erg/s$ at maximum,
%}}

%\noindent \colorbox{yellow}{\parbox{\linewidth}{
\begin{equation}
P_{\gamma} = P_{a \rightarrow \gamma} \cdot \dfrac{(0.5 d_{spot})^2}{(\tan \left( \alpha / 2 \right) \cdot 0.3 R_{Sun})^2} \cdot \Lambda_a \approx 4.9 \cdot
10^{-3}, \label{eq3.35}
\end{equation}
%}}

%\noindent \colorbox{yellow}{\parbox{\linewidth}{
\noindent directly following from the geometry of the system (Fig.~\ref{fig-lampochka}b), where the conversion probability $P_{a \rightarrow \gamma} \sim
0.0114$ (\ref{eq3.31}), $d_{spot} \sim 60 \cdot 10^3 ~km$ is the measured diameter of the sunspot (umbra)~\cite{Dikpati2008,Gough2010}. Its size
determines the relative portion of the axions hitting the sunspot area. Further,

%\noindent \colorbox{yellow}{\parbox{\linewidth}{
\begin{equation}
\dfrac{(0.5 d_{spot})^2}{(\tan \left( \alpha / 2 \right) \cdot 0.3 R_{Sun})^2} \cong 1, \nonumber
\end{equation}
%}}

\noindent and the value $\Lambda_a$ characterizes the portion of the channeling axion flux corresponding to the total $(2\left\langle N_{spot} \right\rangle
_{max})$ sunspots on the photosphere:

%\noindent \colorbox{yellow}{\parbox{\linewidth}{
\begin{equation}
\Lambda_a = \dfrac{\left( channeling\ axion\ flux \right)}{(1/3)\left( total\ axion\ flux \right)} \approx \dfrac{2 \left\langle N_{spot} \right\rangle
_{max}}{(4/3) \left( R_{Sun} / 0.5 d_{spot} \right)^2} \sim 0.42 , \label{eq3.36}
\end{equation}
%}}

\noindent where $\left\langle N_{spot} \right\rangle _{max} \approx 150$ is the average number of the maximal sunspot number (over the visible
hemisphere~\cite{Dikpati2008,Gough2010}) for the cycle 22 experimentally observed by the  Japanese X-ray telescope Yohkoh (1991)~\cite{ref36}.

%\noindent \colorbox{yellow}{\parbox{\linewidth}{
On the other hand, from the known observations (see~\cite{ref59} and Appendix \ref{appendix-luminosity})
%}}

%\noindent \colorbox{yellow}{\parbox{\linewidth}{
\begin{equation}
\frac{(L_{corona}^X)_{max}}{L_{Sun}} \cong 1.22 \cdot 10^{-6},
\label{eq3.37}
\end{equation}
%}}

%\noindent \colorbox{yellow}{\parbox{\linewidth}{
\noindent where $L_{Sun} = 3.8418 \cdot 10^{33} erg / s$ is the solar luminosity~\cite{ref63}. Using the theoretical axion impact estimate (\ref{eq3.32}), one
can clearly see that the obtained value (\ref{eq3.35}) is in good agreement with the observations (\ref{eq3.37}):
%}}

%\noindent \colorbox{yellow}{\parbox{\linewidth}{
\begin{equation}
P_{\gamma} =  \left. \frac{(L_{corona}^X)_{max}}{L_{Sun}} \middle/ \frac{L_a}{L_{Sun}} \sim 4.9 \cdot 10^{-3} \right. ,
\label{eq3.38}
\end{equation}
%}}

%\noindent \colorbox{yellow}{\parbox{\linewidth}{
\noindent derived independently.
%}}

%\noindent \colorbox{yellow}{\parbox{\linewidth}{
In other words, if the hadronic axions found in the Sun are the same particles found in the white dwarfs with the known strength of the axion coupling to
photons (see (\ref{eq3.30})), it is quite natural that the independent observations give the same estimate of the probability $P_{\gamma}$ (see (\ref{eq3.35})
and (\ref{eq3.38})). So the consequences of the choice (\ref{eq3.30}) are determined by the independent measurements of the average sunspot radius, the sunspot
number~\cite{Dikpati2008,Gough2010}, the model estimates of the horizontal magnetic field and the height $L_{MS}$ of the magnetic steps (see Fig.
\ref{fig-lampochka}), and the hard part of the solar photon spectrum mainly determined by the axion-photon conversion efficiency, and the theoretical estimate
for the part of the axion luminosity $L_a$ in the total luminosity of the Sun $L_{Sun}$ (\ref{eq3.38}).

\section{Axion mechanism of the solar Equator -- Poles effect}

\begin{figure*}
%\centerline{\includegraphics[width=12cm]{Sun_X-ray_image_spectrum.eps}}
\centerline{\includegraphics[width=12cm]{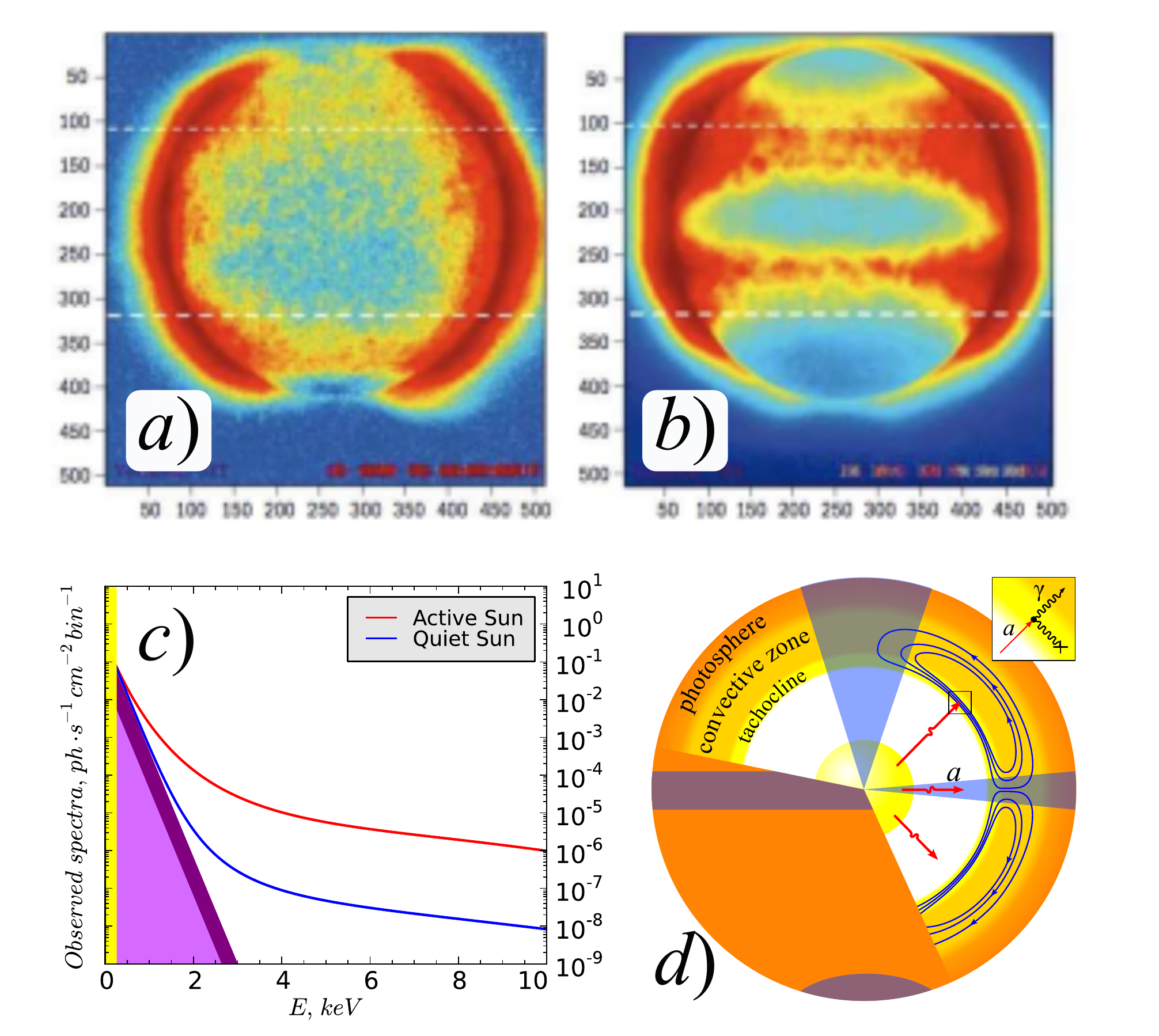}}

\caption{\textbf{Top:} Solar images at photon energies from 250~eV up to a few keV from the Japanese X-ray telescope Yohkoh (1991-2001) (adopted
from~\cite{ref36}). The following is shown: \newline (a) a composite of 49 of the quietest solar periods during the solar minimum in 1996; \newline (b) solar
X-ray activity during the last maximum of the 11-year solar cycle. Most of the X-ray solar activity (right) occurs at a wide bandwidth of $\pm 45^{\circ}$ in
latitude, being homogeneous in longitude. Note that $\sim$95\% of the solar magnetic activity covers this bandwidth. \newline \textbf{Bottom:} (c) Axion
mechanism of solar irradiance variations with the soft part of the solar photon spectrum independent of the Sun's phases (brown band) and the red
and blue curves characterizing the irradiance increment in the active and quiet phases of the Sun, respectively; \newline (d) schematic picture of the radial
travelling  of the axions inside the Sun. Blue lines on the Sun designate the magnetic field. In the tachocline axions are converted into $\gamma$-quanta, that
move towards the poles (blue cones) which form the experimentally observed Solar photon spectrum after passing the photosphere (Fig.~\ref{fig06}). Solar axions
in the equatorial plane (blue bandwidth) are not converted by Primakoff effect (inset: diagram of the inverse coherent process). The variations of the solar
axions may be observed at the Earth by special detectors like the new generation CAST-helioscopes~\cite{ref68}. }
%\label{fig-Yohkoh}
\label{fig-Yohkoh}
\end{figure*}

The axion mechanism of Sun luminosity is largely validated by the experimental X-ray images of the Sun in the quiet (Fig.~\ref{fig-Yohkoh}a) and active
(Fig.~\ref{fig-Yohkoh}b) phases~\cite{ref36} which clearly reveal the so-called Solar Equator -- Poles effect (Fig.~\ref{fig-Yohkoh}b).

The essence of this effect lies in the following. It is known that the axions may be transformed into $\gamma$-quanta by inverse Primakoff effect in the
transverse magnetic field only. Therefore the axions that pass towards the poles (blue cones in Fig.~\ref{fig-Yohkoh}b) and equator (the blue band in
Fig.~\ref{fig-Yohkoh}b) are not transformed into $\gamma$-quanta by inverse Primakoff effect, since the magnetic field vector is almost collinear to the axions'
momentum vector. The observed nontrivial X-ray distribution in the active phase of the Sun may be easily and naturally described within the framework of the
axion mechanism of Sun luminosity.

%\textcolor{red}{---------------------------------}

\begin{figure*}
%\centerline{\includegraphics[width=15cm]{axion-channaling-scattering-Yohkoh-3.eps}}
\centerline{\includegraphics[width=15cm]{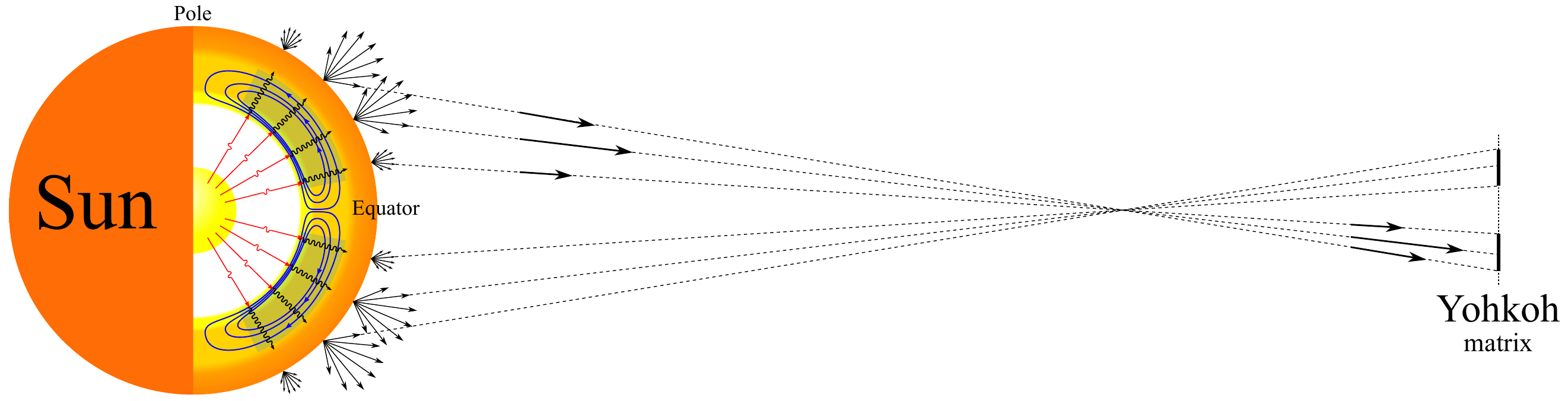}}

\caption{The formation of the high X-ray intensity bands on the Yohkoh
matrix. \label{fig07a}}
\end{figure*}

After the axions transformation into photons in the tachocline, these photons travel through the entire convective zone along the magnetic flux tubes, as
described in Section~\ref{subsec-channeling}, up to the photosphere. In the photosphere these photons undergo a multiple Compton scattering (see
Section~\ref{subsec-osc-parameters}) which results in a substantial deviation from the initial axions directions of propagation (Fig.~\ref{fig07a}).

%\noindent\colorbox{yellow}{\parbox{\linewidth}{

Let us make a simple estimate of the Compton scattering efficiency in terms of
the X-ray photon mean free path (MFP) in the photosphere:

\begin{equation}
l = \left( \sigma_c \cdot n_e \right)^{-1}\, , \label{eq-compt-01}
\end{equation}
where the total Compton cross-section $\sigma_c = \sigma_0 = 8 \pi r_0^2 / 3$ for the low-energy photons \cite{ref81,ref82}, $n_e$ is the electrons density in
the photosphere, and $r_0 = 2.8\cdot10^{-13}~cm$ is the so-called classical electron radius.

%\begin{align}
%\sigma_c = \sigma_0 \cdot \frac{3}{8\varepsilon} \left\lbrace \left[ 1 - \frac{2(\varepsilon + 1)}{\varepsilon^2} \right] &
%\ln ({2\varepsilon + 1}) + \frac{1}{2} \right. +  \nonumber \\
%& \left.  + \frac{4}{\varepsilon} - \frac{1}{2(2\varepsilon + 1)^2} \right\rbrace
%\end{align}

Taking into account the widely used value of the matter density in the solar
photosphere $\rho \sim 10^{-7} ~g/cm^3$ and supposing that it consists of the
hydrogen (for the sake of the estimation only), we obtain that

\begin{equation}
n_e \approx \frac{\rho}{m_H}  \approx 6 \cdot 10^{16} ~ electron / cm^3\, ,
\label{eq-compt-02}
\end{equation}
which yields the MFP of the photon \cite{ref81,ref82}
\begin{align}
l = \left( 7 \cdot 10^{-25} ~cm^2 \cdot 6 \cdot 10^{16} ~electron/cm^3 \right)^{-1} \approx  2.4 \cdot 10^7 ~cm = 2.4 \cdot 10^5 ~m\, . \label{eq-compt-03}
\end{align}

Since this value is smaller than the thickness of the solar photosphere ($\sim 10^{6}~m$), the Compton scattering is efficient enough to be detected at the
Earth (see Fig.~\ref{fig07a} and Fig.~10 in~\cite{ref36}).

%}}

Taking into account the directional patterns of the resulting radiation as well as the fact that the maximum of the axion-originated X-ray radiation is
situated near 30 -- 40 degrees of latitude (because of the solar magnetic field configuration), the mechanism of the high X-ray intensity bands formation on
the Yohkoh matrix becomes obvious. The effect of these bands widening near the edges of the image is discussed in Appendix~\ref{appendix-widening} in detail.

\section{Summary and Conclusions}

In the given paper we present a self-consistent model of the axion mechanism of Sun luminosity, in the framework of which we estimate the values of the axion
mass ($m_a \sim 2.3 \cdot 10^{-2} ~eV$) and the axion coupling constant to photons ($g_{a \gamma} \sim 3.6 \cdot 10^{-11} ~GeV^{-1}$). A good correspondence
between the solar axion-photon oscillation parameters and the hadron axion-photon coupling derived from white dwarf cooling (see Fig.~\ref{fig05}) is
demonstrated.

%It is necessary to note that obtained estimations can't be excluded by the existing experimental data because the discussed above effect of solar axion
%intensity modulation by temporal variations of the toroidal magnetic field of the solar tachocline zone was not taken into account in these observations. On
%the other hand, the obtained estimates for the axion-photon coupling cannot be ruled out by the existing theoretical limitations known as the globular cluster
%star limit ($g_{a \gamma} < 6 \cdot 10^{-11} ~GeV^{-1}$), since these values are highly model-dependent. It actually means that the axion parameters obtained
%in the present paper do not contradict any of the known experimental and theoretical model-independent limitations.

Below we give some major ideas of the paper which formed a basis for the statement of the problem justification, and the corresponding experimental data which
comport with these ideas either explicitly or implicitly.

%\subsection{Axion mechanism of Sun luminosity}

One of the key ideas behind the axion mechanism of Sun luminosity is the effect of $\gamma$-quanta channeling along the magnetic flux tubes (waveguides inside
the cool region) in the Sun convective zone (Figs.~\ref{fig04} and~\ref{fig-lampochka}). The low refraction (i.e. the high transparency) of the thin magnetic
flux tubes is achieved due to the ultrahigh magnetic pressure (see (\ref{eq37})), induced by the magnetic field of about 2000~T (Fig.~\ref{fig03}a). So it may
be concluded that the axion mechanism of Sun luminosity based on the lossless $\gamma$-quanta channeling along the magnetic tubes allows to explain the effect
of the partial suppression of the convective heat transfer, and thus to understand the known puzzling darkness of the sunspots~\cite{ref34-3}.

\begin{figure}
%\noindent \colorbox{yellow}{\parbox{\linewidth}{
\begin{center}
\includegraphics[width=15.5cm]{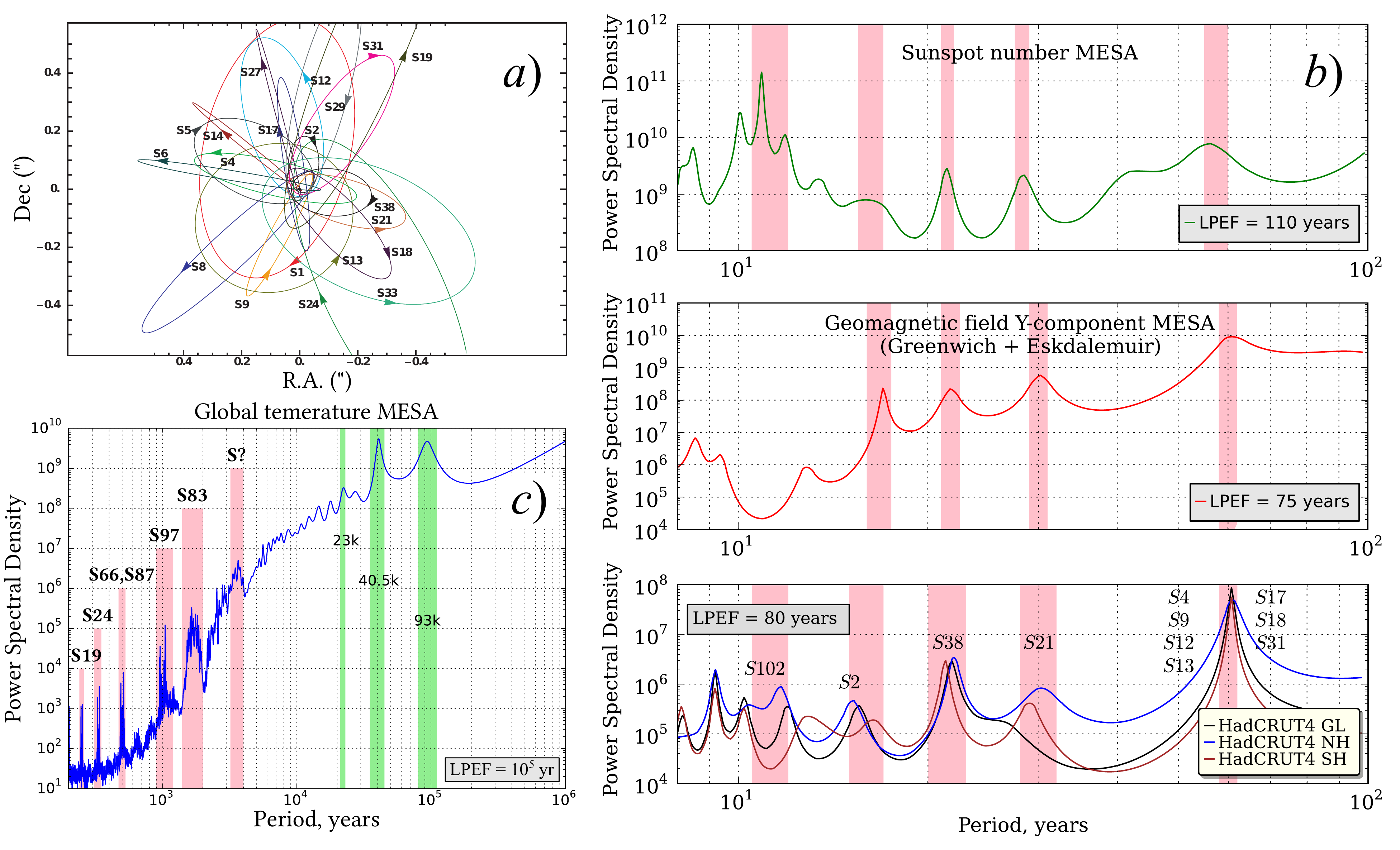}
\end{center}
%}}
\caption{a) Stellar orbits at the Galactic Center in the central arcsecond
         (declination Dec(")  as a function of time for the stars and red
         ascension R.A.("))~\cite{Gillessen2009,Genzel2010}. The coordinate system
         is chosen so that Sgr A* (the supermassive black hole with the mass of
         $\sim$4.3$\cdot 10^6$~M$_{\odot}$) is at rest.\newline
         b) Power spectra of the sunspots (1874-2010 from Royal Greenwich
         Observatory), Geomagnetic field Y-component (from Greenwich and
         Eskdalemuir~\cite{Edinburg} observatories) and HadCRUT4 GST
         (1850-2012) (black), Northern Hemisphere (NH) and Southern Hemisphere
         (SN) using the maximum entropy spectral analysis (MESA); red boxes
         represent major astronomical oscillations associated to the major
         heliospheric harmonics associated to the orbits of the best known
         short-period S-stars (S0-102, S2, S38, S21, S4-S9-S12-S13-S17-S18-S31)
         at the Galactic Center~\cite{Gillessen2009,Genzel2010,Gillessen2013}
         and to the solar cycles (about 11-12, 15-16, 20-22, 29-30, 60-61 years).\newline
         c) Power spectra of the global temperature as reconstructed
         in~\cite{Bintanja2008}; red boxes represent the major astronomical
         oscillations associated to the major heliospheric harmonics associated
         to the orbits of the best known non-short-period S-stars (S19, S24,
         S66-S87, S97, S83, S?) at the Galactic
         Center~\cite{Gillessen2009,Genzel2010} and to the solar cycles (about 250,
         330, 500, 1050, 1700, 3600 years); green boxes represent the major
         temperature oscillations presumably associated with the variations of
         the Earth orbital parameters: eccentricity ($\sim$93~kyr), obliquity
         ($\sim$41~kyr) and axis precession ($\sim$23~kyr).}
\label{fig-MESA}
\end{figure}

%\noindent \colorbox{yellow}{\parbox{\linewidth}{
From here it follows as a supposition that the physics of the connection "local (magnetic?) shear -- strong cooling -- axion mechanism of $\gamma$-quanta
channeling" is completely based on the processes of the local shear formation and variations, which is still not studied. We also suggest a possible mechanism
of the azimuthal magnetic flux formation and concentration near the overshoot tachocline to be driven by the interaction of the nuclei within the Sun with the
asymmetric dark matter (see~\cite{Frandsen2010}), the flow intensity of which may depend on the stellar orbits around the massive black hole in the center of
the Milky Way (see~\cite{Gillessen2009,Meyer2014}). The further development of this idea is a subject of our forthcoming paper. It is important to note that
the basis for this is the experimental data on the 11-year sunspot number oscillation (Fig.~\ref{fig-MESA}b) and the geophysical parameters
(Fig.\ref{fig-MESA}b,c). In our view, all these parameters may be determined by the motion of the short orbiting stars around the black hole at the center of
our galaxy (see Fig.~\ref{fig-MESA}a~\cite{Gillessen2009,Meyer2014}).
%}}

It is shown that the axion mechanism of luminosity variations (which means that they are produced by adding the intensity variations of the $\gamma$-quanta of
the axion origin to the invariant part of the solar photon spectrum (Fig.~\ref{fig06}b)) easily explains the physics of the so-called Solar Equator -- Poles
effect observed in the form of the anomalous X-ray distribution over the surface of the active Sun, recorded by the Japanese X-ray telescope Yohkoh
(Fig.~\ref{fig-Yohkoh}, top).

The essence of this effect consists in the following: axions that move towards the poles (blue cones in Fig.~\ref{fig-Yohkoh}, bottom) and equator (blue
bandwidth in Fig.~\ref{fig-Yohkoh}, bottom) are not transformed into $\gamma$-quanta by the inverse Primakoff effect, because the magnetic field vector is
almost collinear to the axions' momentum in these regions (see the inset in Fig.~\ref{fig-Yohkoh}, bottom). Therefore the anomalous X-ray distribution over the
surface of the active Sun is a kind of a "photo" of the regions where the axions' momentum is orthogonal to the magnetic field vector in the solar over-shoot
tachocline. The solar Equator -- Poles effect is not observed during the quiet phase of the Sun because of the magnetic field weakness in the overshoot
tachocline, since the luminosity increment of the axion origin is extremely small in the quiet phase as compared to the active phase of the Sun.

In this sense, the experimental observation of the solar Equator -- Poles effect is the most striking evidence of the axion mechanism of Sun luminosity
variations. It is hard to imagine another model or considerations which would explain such anomalous X-ray radiation distribution over the active Sun surface
just as well (compare Fig.~\ref{fig-Yohkoh}a,b with Fig.~\ref{app-b-fig01}a and Fig.~\ref{fig-MESA}b,c).

%\

%\subsection{Invisible axions and solar Equator -- Poles effect}
%
%Obviously, if the axion mechanism of Sun luminosity exists, the new limitations on the axion-photon coupling must be calculated for the future experiments and
%re-calculated for the old ones taking into account the solar equator effect and the contribution of the inverse Primakoff effect into the total Sun luminosity.
%This conclusion is extremely important, since the exact value of the axion-photon coupling plays a major role in the particle physics, astrophysics and
%cosmology.

And, finally, let us emphasize one essential and the most painful point of the present paper. It is related to the key problem of the axion mechanism of Sun
luminosity and is stated rather simply: "Is the process of axion conversion into $\gamma$-quanta by the Primakoff effect really possible
%in the Solar tachocline magnetic field?"
in the magnetic steps of an O-loop near the solar overshoot tachocline?" This question is directly connected to the problem of the hollow magnetic flux tubes
existence in the convective zone of the Sun, which are supposed to connect the tachocline with the photosphere. So, either the more general theory of the Sun
or the experiment have to answer the question of whether there are the waveguides in the form of the hollow magnetic flux tubes in the cool region of the
convective zone of the Sun, which are perfectly transparent for $\gamma$-quanta, or our model of the axion mechanism of Sun luminosity is built around simply
guessed rules of calculation which do not reflect any real nature of things.

\

\section*{Acknowledgements}

\noindent The work of V.D.~Rusov, T.N.~Zelentsova and V.P.~Smolyar is partially supported by EU FP7 Marie Curie Actions, SP3-People, IRSES project
BlackSeaHazNet (PIRSES-GA-2009-246874).

\noindent The work of M. Eingorn was supported by NSF CREST award HRD-1345219 and NASA grant NNX09AV07A.

\appendix
\numberwithin{equation}{section}
\numberwithin{figure}{section}
\numberwithin{table}{section}

\

%\noindent \colorbox{yellow}{\parbox{\linewidth}{
\section{X-ray coronal luminosity variations}
\label{appendix-luminosity}
%}}

%\noindent\colorbox{yellow}{\parbox{\linewidth}{

\begin{figure*}[htb!]
\noindent
\begin{center}
\includegraphics[width=15.5cm]{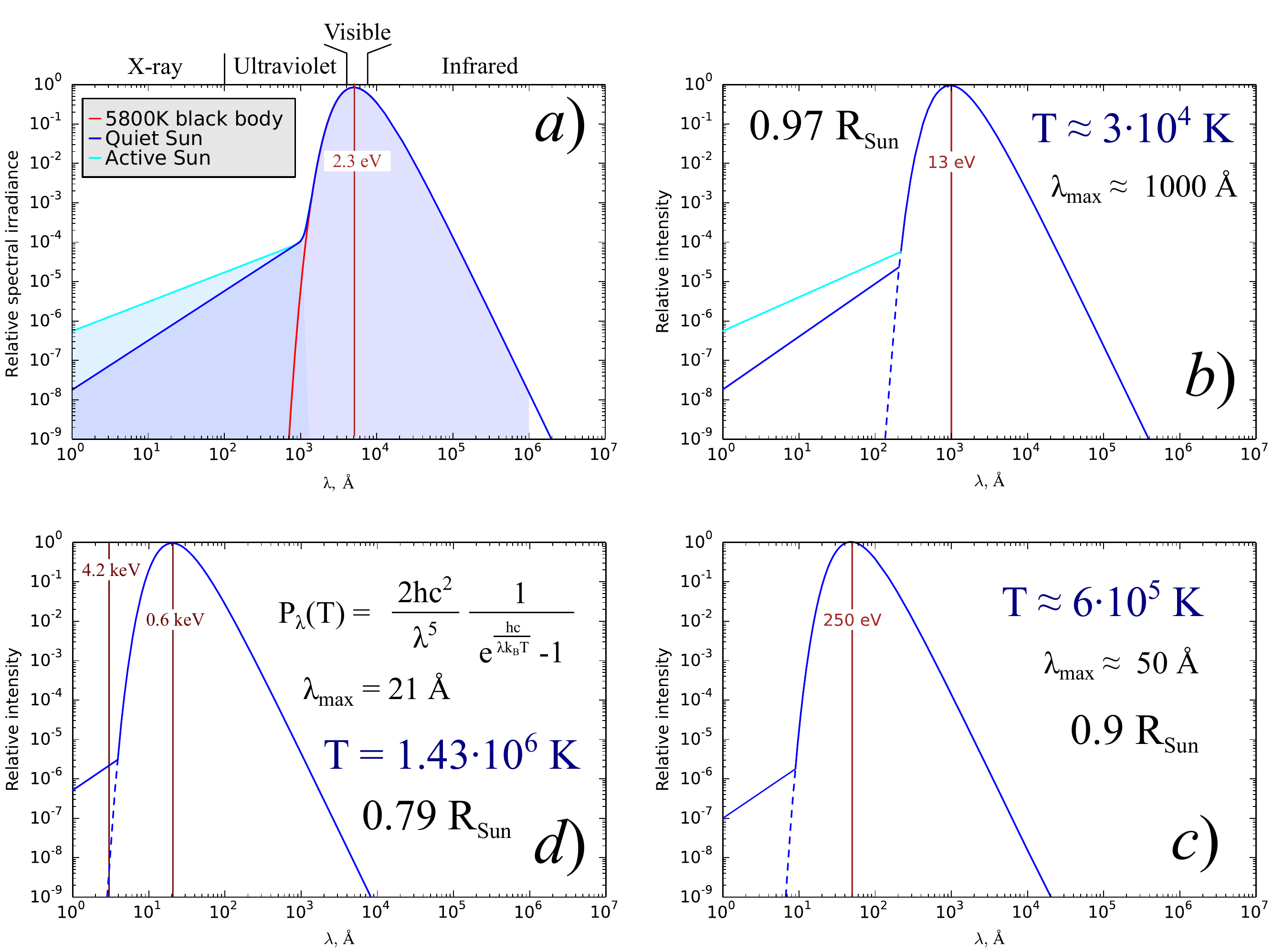}
\end{center}
%\centerline{\includegraphics[width=16cm]{sun_approx_spectrum-6.pdf}}
\caption{a) The smoothed Solar spectrum corresponds to a black body with a
         temperature $5.77 \cdot 10^3$~K at $R_{Sun}$ (see Fig.~11
         in~\cite{ref36}). The light cyan shaded area marks the band considered
         the "X-ray" band in our calculations, and the blue shaded area
         corresponds to the "Optical" band.\newline
         b) A spectrum of a black body with a temperature $3 \cdot 10^4$~K
         at 0.97$R_{Sun}$~\cite{ref44-3,ref45-3}. Using a detailed numerical
         description of the solar interior of the standard model with helium
         diffusion (see Table XVI in~\cite{ref45-3}) it may be shown that the
         X-ray luminosity depends on Compton scattering of the photons outside
         the magnetic tubes (blue line) and the "channeled" photons of the
         axion origin (cyan line).\newline
         c) A spectrum of a black body with a temperature $6 \cdot 10^5$~K
         at 0.9$R_{Sun}$~\cite{ref44-3,ref45-3}. X-ray luminosity (blue line)
         is determined by the Compton scattering of the photons outside the
         magnetic tubes only, since the "channeled" axion-originated photons
         almost do not experience the Compton scattering inside the magnetic
         tubes.\newline
         d) A spectrum of a black body with a temperature
         $1.43 \cdot 10^6$~K at 0.79$R_{Sun}$~\cite{ref44-3,ref45-3}. X-ray
         luminosity (blue line) is determined by the Compton scattering of the
         photons outside the magnetic tubes only, since there is certainly no
         scattering in the cool region of the magnetic tubes.}
\label{app-b-fig01}
\end{figure*}

The X-ray luminosity of the Solar corona during the active phase of the solar cycle is defined by the following expression:
\begin{align}
\left( L_{corona} ^{X} \right)_{max} = \int \limits_{X-ray} \frac{d\Phi_{corona}^{max} (E)}{dE} \cdot E dE =  \int
\limits_{X-ray} \frac{d W_{corona}^{max} (E)}{dE} dE\, .
\label{app-b-eq01}
\end{align}

In the quiet phase it may be written as
\begin{align}
\left( L_{corona} ^{X} \right)_{min} = \int \limits_{X-ray} \frac{d\Phi_{corona}^{min} (E)}{dE} \cdot E dE =  \int
\limits_{X-ray} \frac{d W_{corona}^{min} (E)}{dE} dE\, .
\label{app-b-eq02}
\end{align}

Then integrating the blue curve in Fig.~\ref{app-b-fig01} for $\left(L_{corona}^X \right) _{max}$ and the cyan curve for $\left(L_{corona}^X \right)_{min}$, we
obtain
%\begin{align}
%\frac{\Delta L_{corona}^X}{L_{Sun}} \sim 10^{-6}\, .
%\end{align}

%}}

\begin{equation}
\frac{\left( L_{corona}^X \right) _{min}}{L_{Sun}} \sim 10^{-7} ; ~~~~~
\frac{\left( L_{corona}^X \right) _{max}}{L_{Sun}} \sim 10^{-6} .
\label{app-b-eq03}
\end{equation}

So it may be derived from here that the Sun's luminosity is quite low in X-rays (\ref{app-b-eq03}), typically (see~\cite{Rieutord2014}),

\begin{equation}
10^{-7} L_{Sun} \leqslant L_{corona}^X \leqslant 10^{-6} L_{Sun},
\label{app-b-eq04}
\end{equation}

\noindent but it varies with the cycle (see blue and cyan lines in Fig.~\ref{app-b-fig01}a) as nicely shown by the pictures obtained with the Yohkoh
satellite (see Fig.~4 in~\cite{Rieutord2014}).

And finally, it may be supposed that X-rays, propagating from the tachocline towards the photosphere, interact with the charged particles via the Compton
scattering, but only outside the magnetic tubes. The axion-originated X-ray radiation channeling inside the magnetic tubes does not experience the Compton
scattering up to the photosphere (Figure~\ref{app-b-fig01}).

\section{Explanation of the high X-ray intensity bands widening near the Yohkoh image edges}
\label{appendix-widening}

It is interesting to note that the bands of high X-ray intensity on Yohkoh images deviate from the solar parallels (Fig.~\ref{fig-Yohkoh}b). This is especially
the case near the edges of the visible solar disk.

\begin{figure*}
%\centerline{\includegraphics[width=15cm]{Sun-Yohkoh.eps}}
\centerline{\includegraphics[width=15cm]{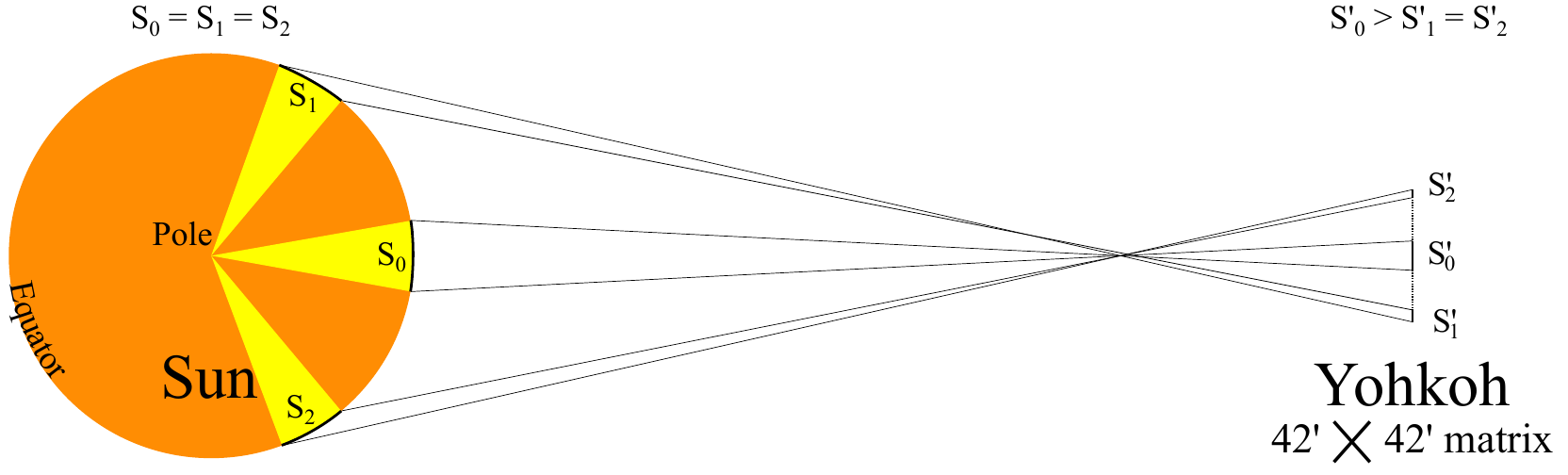}}

\caption{A sketch of the Sun image formation on the Yohkoh matrix. \label{app-a-fig05}}
\end{figure*}

\begin{figure*}
%\centerline{\includegraphics[width=9cm]{Sun-meridians-Yohkoh.eps}}
\centerline{\includegraphics[width=9cm]{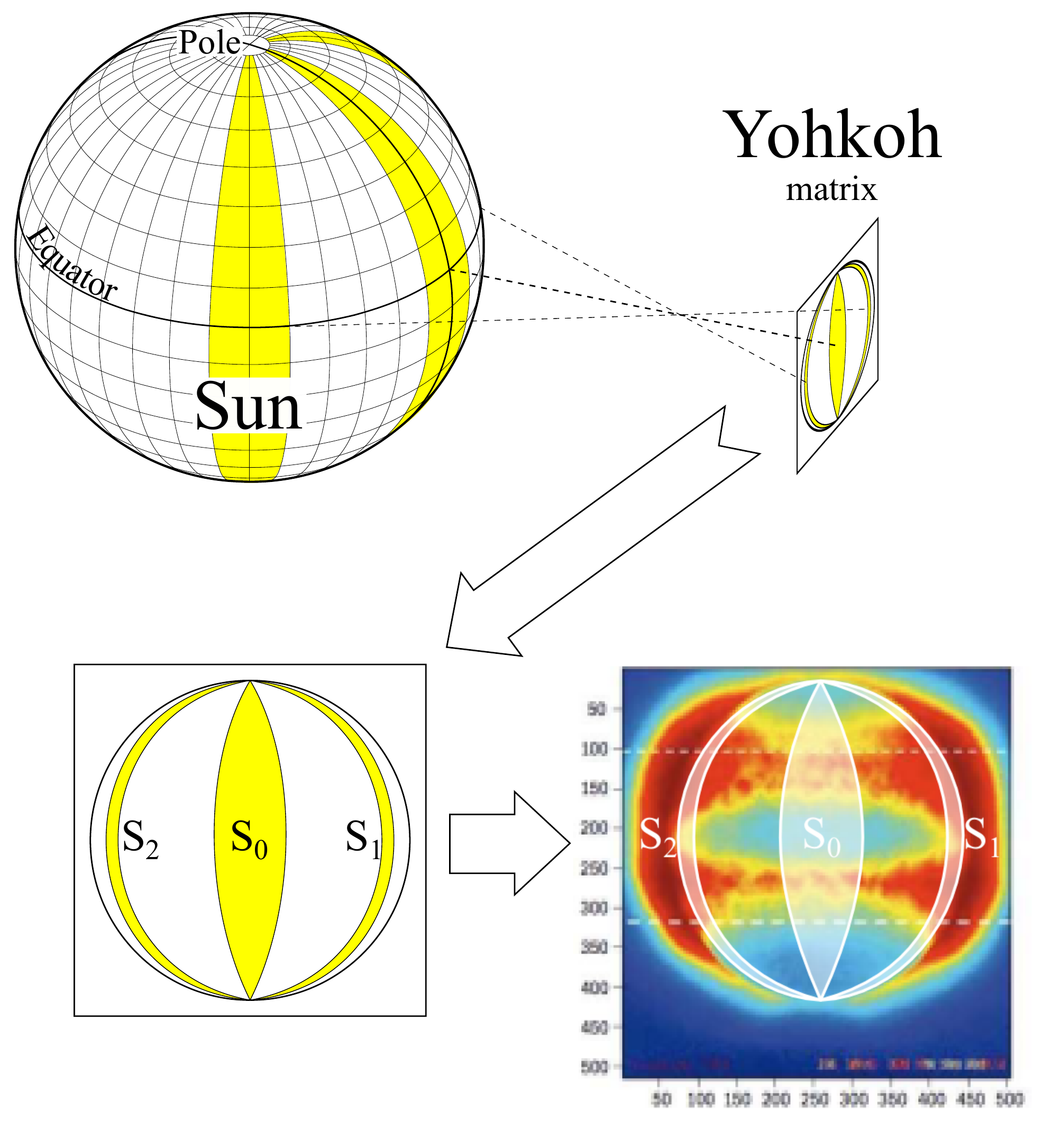}}

\caption{A sketch of the Sun image formation on the Yohkoh matrix. \label{app-a-fig04}}
\end{figure*}

This effect may be explained graphically by means of figures~\ref{app-a-fig05} and~\ref{app-a-fig04}. These figures show the schematic concept of the Sun image
formation on the Yohkoh matrix. Fig.~\ref{app-a-fig05} shows the Sun from its pole.

Let us choose three sectors of equal size on the surface of the Sun (Fig.~\ref{app-a-fig05}, \ref{app-a-fig04}). The areas of the photosphere cut by these
sectors are also equal ($S_0 = S_1 = S_2$). However, as it is easily seen in the suggested scheme, the \emph{projections} of these sectors on the Yohkoh matrix
are not of equal area (Fig.~\ref{app-a-fig05}, \ref{app-a-fig04}). The $S'_1$ and $S'_2$ projections areas are much less than that of the $S'_0$ projection
(Fig.~\ref{app-a-fig05}). It means that the radiation emitted by the sectors $S_1$ and $S_2$ of the solar photosphere and captured by the satellite camera will
be concentrated within \emph{less} area (near the edges of the solar disk) than the radiation coming from the $S_0$ sector (in the center of the solar disk).
As a result, the satellite shows higher intensity near the image edges than that in the center, in spite of the obvious fact that the real radiation intensity
is equal along the parallel of the Sun.

Therefore, because of the system geometry, the satellite tends to ``amplify'' the intensity near the image edges, and the areas that correspond to the yellow
and green areas at the center (Fig.~\ref{fig-Yohkoh}b) become red near the edges, thus leading to a visible widening of the high intensity bands.

A particularly high radiation intensity near the very edges of the visible solar disk, observed even during the quiet phase of the Sun (Fig.~\ref{fig-Yohkoh}a),
indicates a rather ``wide'' directional radiation pattern of the solar X-rays.

%\subsection{Simulation}

Let us make a simple computational experiment. We will choose a sphere of a unit radius and spread the points over its surface in such a way that their density
changes smoothly according to some dependence of the \emph{polar} angle ($\theta$). The \emph{azimuth} angle will not influence the density of these points.
For this purpose any function that provides a smooth change of the density will do. For example, this one:
\begin{equation}
\rho (\theta) = [\rho _0 + \rho _{max} \cdot \cos(2 \cos\theta)]^{-1}\, . \label{app-a-eq01}
\end{equation}

\begin{figure*}
%\centerline{\includegraphics[width=9cm]{points_density_function.eps}}
\centerline{\includegraphics[width=9cm]{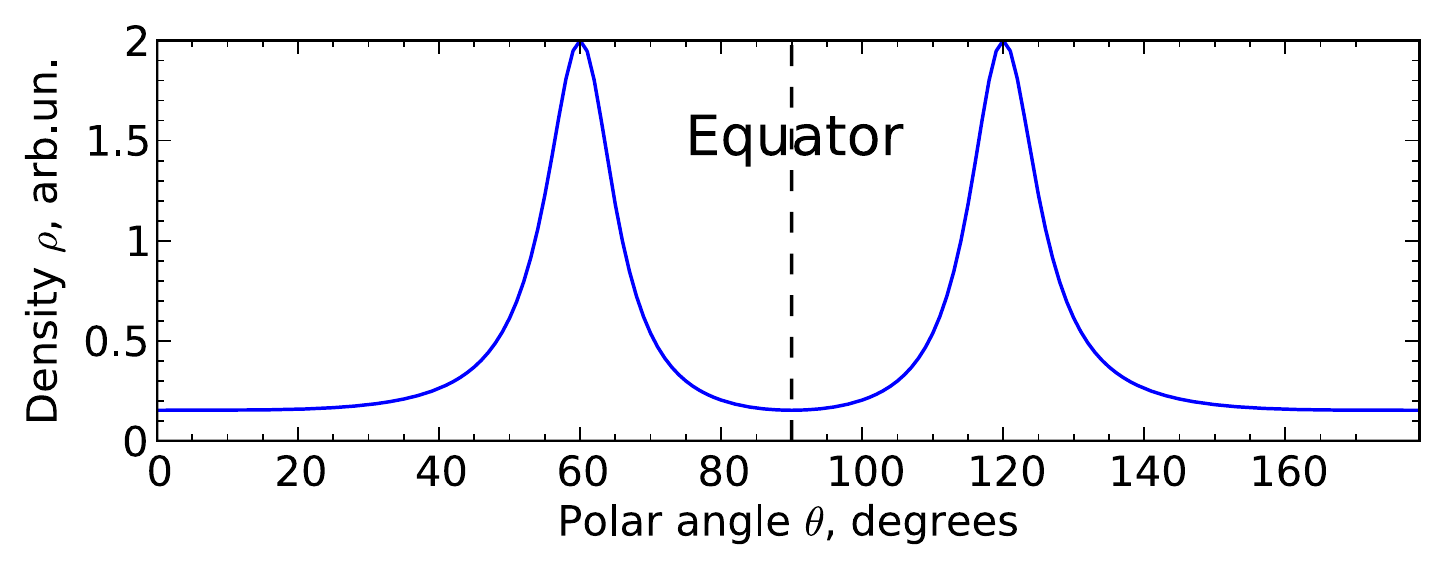}}

\caption{Graphical representation of Eq.~(\ref{app-a-eq01}). \label{app-a-fig06}}
\end{figure*}

Here we take $\rho_0=3.5$ and $\rho_{max}=3$ in arbitrary units. The graphical representation of this dependence is shown in Fig.~\ref{app-a-fig06}.

I.e. it yields a minimum density of the points near the poles and the equator, and the maximum density of the points near $\theta = 60^{\circ}$ and $\theta =
120^{\circ}$.

The polar angle $\theta$ was set in the range $[0, 180^{\circ}]$ with the step of $1^{\circ}$. The azimuth angle $\varphi$ was set in the range $[0,
180^{\circ}]$ (one hemisphere) with a variable step $\Delta \varphi$ representing the variable density, since the points density is inversely proportional to
the step between them ($\Delta \varphi \sim 1 / \rho$). We assume that
\begin{equation}
\Delta \varphi (\theta) = \Delta \varphi _0 + \Delta \varphi _{max} \cdot \cos(2 \cos\theta) ~~[deg]\, . \label{app-a-eq01a}
\end{equation}

The values of $\Delta \varphi _0 = 3.5^{\circ}$ and $\Delta \varphi _{max} = 3^{\circ}$ were chosen arbitrarily. So,
\begin{align}
\Delta \varphi (\theta) = 3.5 + 3 \cdot \cos(2 \cos\theta) ~~[deg]\, . \label{app-a-eq03}
\end{align}

From Eq.~(\ref{app-a-eq03}) it is clear that the minimum step was 0.5$^{\circ}$ and the maximum step was 6.5$^{\circ}$. Apparently, the more is the step, the
less is the density (near the poles and the equator) and vice versa, the less is the step, the more is the points density. This was the way of providing a
smooth change of the points density by latitude (along the solar meridians).

\begin{figure*}
  \begin{center}
  \begin{minipage}[h]{0.49\linewidth}
%  \center{a) \includegraphics[width=5cm]{Sun-bands-3.eps}}
    \center{a) \includegraphics[width=5cm]{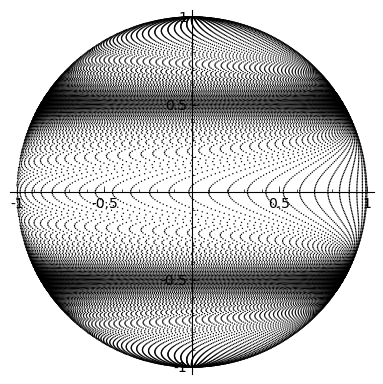}}
  \end{minipage}
  \begin{minipage}[h]{0.49\linewidth}
%  \center{b) \includegraphics[width=6cm]{Sun_X-ray_image-active.eps}}
    \center{b) \includegraphics[width=6cm]{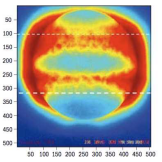}}
  \end{minipage}
  \caption{a) Simulation of the high intensity bands formation on the 2D projection of the sphere.
  \newline b) Sun X-ray image from Yohkoh satellite during the active
  phase of the Sun.}
  \label{app-a-fig07}
  \end{center}
\end{figure*}

Obviously, this forms the "belts" of high density of the points along the parallels. The projection of such sphere on any plane perpendicular to its equator
plane will have the form shown in Fig.~\ref{app-a-fig07}a. As it is seen in this figure, although the density does not depend on azimuth angle, there is a high
density bands widening near the edges of the \emph{projected} image. These bands are similar to those observed on the images of the Sun in
Fig.~\ref{app-a-fig07}b.

%\begin{figure}[htb!]
%  \begin{center}
%  \includegraphics[width=10cm]{Sun-bands-3.png}
%  \caption{Simulation.}
%  \label{fig06}
%  \end{center}
%\end{figure}

Let us emphasize that the exact form of the dependence (\ref{app-a-eq01a}) as well as the exact values of its parameters were chosen absolutely arbitrarily for
the sole purpose of the qualitative effect demonstration. They have no relation to the actual latitudinal X-ray intensity distribution over the surface of the
Sun.

\bibliography{Rusov-AxionSunLuminosity}

\end{document}